\documentclass[10pt]{article}
\usepackage{euscript,amsmath,amssymb,amsfonts,graphicx,bm,cite}
\usepackage[justification=justified,format=plain]{caption}
  \usepackage{paralist}
  \usepackage{epstopdf}
  \usepackage{graphics} 
  \usepackage{enumitem}
 \usepackage[colorlinks=true]{hyperref}
 \hypersetup{urlcolor=blue, citecolor=red}
 \usepackage{xcolor}
 
\usepackage{changepage}

\usepackage[utf8x]{inputenc}

\usepackage{textcomp,marvosym}
 
\newcommand{\mc}{\mathcal}
\newcommand{\pd}{\partial}

\newcommand{\e}{{\rm e}}

\renewcommand{\d}{{\rm d}}

\newcommand{\bu}{\bar{u}}

\newcommand{\balpha}{\bar{\alpha}}

\usepackage[aboveskip=1pt,labelfont=bf,labelsep=period,justification=raggedright,singlelinecheck=off]{caption}

\date{}

\usepackage{lastpage,fancyhdr,graphicx}
\usepackage{epstopdf}

\setlist[itemize]{leftmargin=*}

\begin{document}
\vspace*{0.2in}

\begin{flushleft}

{\Large \bf Social inhibition maintains adaptivity and consensus of foraging honeybee swarms in dynamic environments}
\newline
\\
\bigskip
Subekshya Bidari\textsuperscript{1}, Orit Peleg\textsuperscript{2}, and Zachary P Kilpatrick\textsuperscript{1,*} 
\\
\bigskip
\textbf{1} Department of Applied Mathematics, University of Colorado, Boulder CO, USA
\\
\textbf{2} Department of Computer Science and BioFrontiers Institute, University of Colorado, Boulder, CO, USA
\\
\textbf{*} zpkilpat@colorado.edu
\end{flushleft}

\section*{Abstract}
To effectively forage in natural environments, organisms must adapt to changes in the quality and yield of food sources across multiple timescales.
Individuals foraging in groups act based on both their private observations and the opinions of their neighbors. How do these information sources interact in changing environments? We address this problem in the context of honeybee swarms, showing inhibitory social interactions help maintain adaptivity and consensus needed for effective foraging. Individual and social interactions of a mathematical swarm model shape the nutrition yield of a group foraging from feeders with temporally switching food quality.
Social interactions improve foraging from a single feeder if temporal switching is fast or feeder quality is low. When the swarm chooses from multiple feeders, the most effective form of social interaction is direct switching, whereby
bees flip the opinion of nestmates foraging at lower yielding feeders.
Model linearization shows that effective social interactions increase the fraction of the swarm at the correct feeder (consensus) and the rate at which bees reach that feeder (adaptivity). Our mathematical framework allows us to compare a suite of social inhibition mechanisms, suggesting experimental protocols for revealing effective swarm foraging strategies in dynamic environments. \\
\vspace{-2mm}

\noindent
{\bf Keywords:} collective decision-making, foraging, optimality, social insects, dynamic environments


\section{Introduction} 
Social insects forage in groups, scouting food sources and sharing information with their neighbors~\cite{sumpter03,visscher07,holldobler08}. The emergent global perspective of animal collectives helps them adapt to dynamic and competitive environments in which food sources' quality and location can vary~\cite{ward08}. Importantly, decisions made by groups involve nonlinear interactions between individuals, temporally integrating information received from neighbors~\cite{ame06}. For example, honeybees {\em waggle dance}\footnote{Worker bees perform this figure-eight dance after returning to the hive from foraging, indicating the direction and distance to water, high-quality flowers, or potential nest sites.~\cite{seeley10}.} to inform nestmates of profitable nectar sources~\cite{seeley00,seeley10}, and use {\em stop signaling}~\footnote{Bees direct a high frequency body vibration at waggle dancers to try and make them stop when problems with nest or feeding sites are detected.~\cite{nieh10}} to dissuade them from perilous food sources~\cite{nieh10} or less suitable nest sites~\cite{seeley12}.
While waggle dancing rouses bees from indecision, stop signaling prevents decision deadlock and builds consensus when two choices are of similar quality~\cite{pais13}. Thus, both positive and negative feedback interactions within the group regulate swarm decisions and foraging~\cite{cinquin02,garnier07}.

Honeybee colonies live in dynamic environments, in which the best adjacent nest or foraging sites can vary across time~\cite{real88,fewell99}. Bees adapt to change by abandoning less profitable nectar sources for those with higher yields~\cite{seeley91}, and by modifying the number of foragers~\cite{benshahar02,tenczar14}. Prior studies focused on how waggle dance recruitment or the heterogeneity of individual bee roles shape swarm adaptivity~\cite{dornhaus04,granovskiy12}. Inhibitory social interactions, whereby bees stop each other from foraging, have been mostly overlooked as a swarm communication mechanism for facilitating adaptation to change~\cite{kietzman15,gray18}. Bayesian principles and experiments suggest individuals discount prior evidence at a timescale matched to the change rate of their environment~\cite{mcnamara06,glaze15}. However, the mechanics of evidence discounting by collectives is not well understood. Here we propose that inhibitory social interactions play an important role in adapting collective beliefs of swarms foraging in a fluid world.

To study how social inhibition shapes foraging yields, we focus on a task in which the nectar quality of feeders is switched periodically. In prior studies~\cite{seeley91,granovskiy12}, swarm foraging targets shifted in response to food quality switches, suggesting bee collectives can detect such changes. Granovskiy et al (2012) emphasized that uncommitted inspector bees  can lead bees away from the previously dominant foraging site~\cite{granovskiy12}. They also found recruitment via waggle dancing is unimportant for effective foraging in changing environments (See also \cite{price19}). Here we also find recruitment can be detrimental, but negative feedback interactions can rapidly pull bees from low to high yielding feeders. This paired with `abandonment' whereby bees spontaneously stop foraging facilitates the swarm-wide temporal discounting of prior evidence. In contrast, strong positive feedback via recruitment causes bees to congregate at feeders even after food quality has dropped, biasing a swarm's behavior based on past states of the world.

We quantify the contribution of these positive and negative feedback interactions within a mathematical swarm model. Our study focuses on four potential inhibitory social interactions --  discriminate and indiscriminate stop signaling~\cite{nieh10,seeley12}, direct switching~\cite{britton02,marshall09}, and self inhibition -- by which foraging bees alter the behavior of other foraging bees.
Strategies are compared by measuring the rate of foraging yield over the timescale of feeder quality switches. When bees have a single feeder, social interactions are less important, but in the case of two feeders the performance of different forms of social interactions is clearly delineated. Direct switching, by which a bee converts another forager to their own preference, is the most effective means for a swarm to adapt to food site quality changes. Also, foraging yields are most sensitive to swarm interaction tuning in rapidly changing environments with lower food quality. Model linearizations allow us calculate a correspondence between social interaction parameters and the {\em consensus} (steady state fraction of bees at the high yielding site) and {\em adaptivity} (the rate of switching from low to high yielding sites). This provides a clear means  of determining the impact of social interactions on a swarm's foraging efficacy.

\section{Results}

The mathematical swarm model assumes individual bees may be uncommitted or committed to one of the possible feeders~\cite{marshall09}. Uncommitted bees spontaneously commit by observing a feeder or by being recruited by another currently foraging bee. Committed bees may spontaneously abandon their chosen feeder, or may be influenced to stop foraging or switch their foraging target based on inhibitory social interactions we describe~\cite{marshall09,seeley12}. A population level model emerges in limit of large swarms. Stochastic effects of the finite system do not qualitatively change our results in most cases (See Appendix~\ref{stochsys}).

\begin{figure}[t]
\begin{center} \includegraphics[width = 14.5cm]{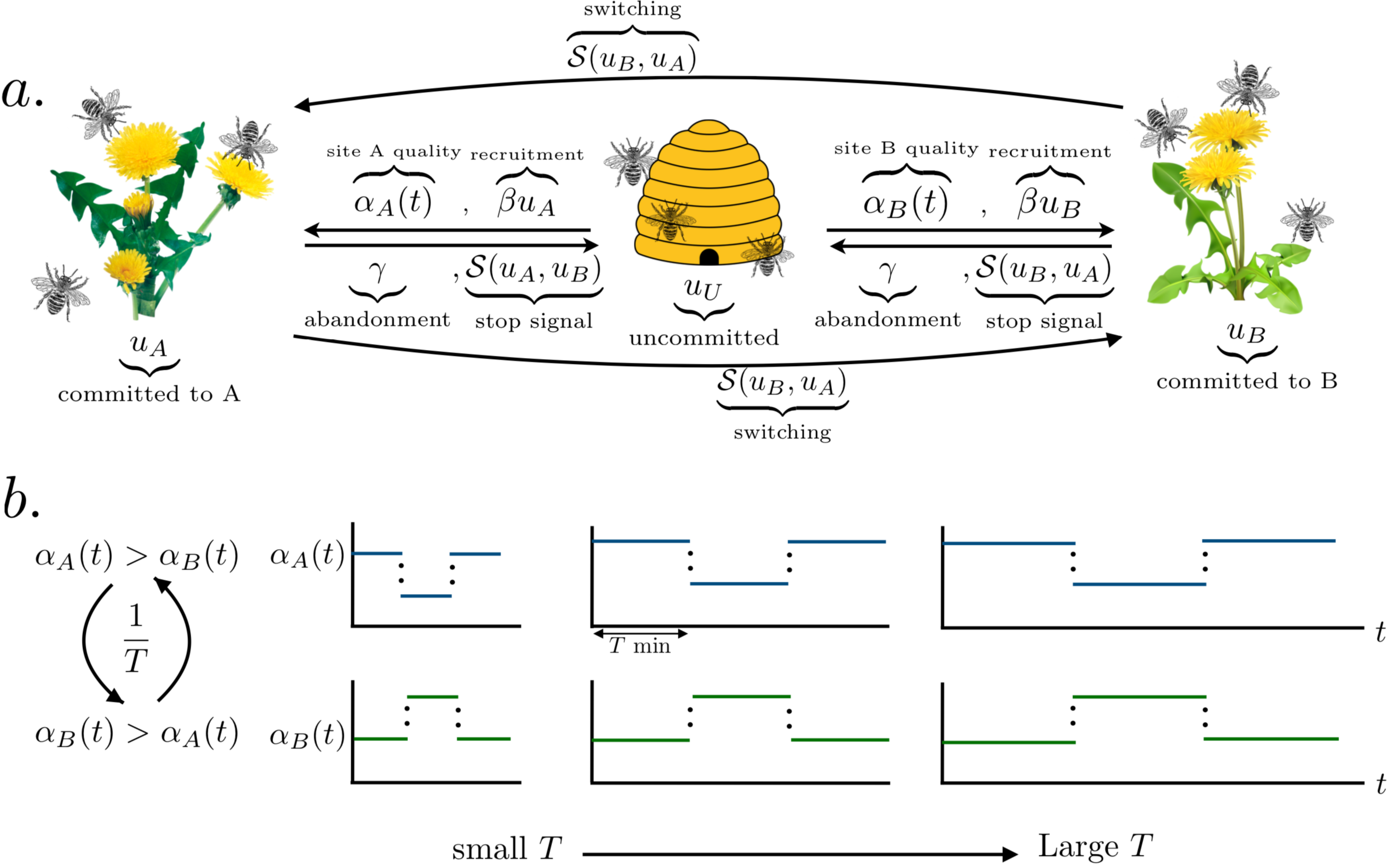} \end{center}
\caption{({\bf a}) Schematic of swarm foraging model with two feeding sites (e.g., flowers or feeder boxes), Eq.~(\ref{2siteswarm}). Bees move along arrows between different opinions (uncommitted or committed); arrow labels indicate interactions that provoke those opinion switches. ({\bf b}) Example feeder quality time series $\alpha_{A,B}(t)$, which switch with period $T$ minutes.}
\label{fig1}
\end{figure}

We mostly focus on two feeder ($A$ and $B$) systems, in which the fraction of the swarm committed to either site is described by a pair of nonlinear differential equations in the limit of large swarms (See Fig.~\ref{fig1}a for a schematic):
\begin{subequations} \label{2siteswarm}
\begin{align}
\dot{u}_A &= (1-u_A - u_B)( \alpha_A(t) + \beta u_A) - \gamma u_A - {\mc S}(u_A, u_B), \\
\dot{u}_B &= (1-u_A - u_B)( \alpha_B(t) + \beta u_B) - \gamma u_B - {\mc S}(u_B, u_A), 
\end{align}
\end{subequations}
where $\alpha_{A,B}(t)$ are time-dependent food qualities at sites $A,B$ (See Fig.~\ref{fig1}b for examples); $\beta$ min$^{-1}$ is the rate bees recruit nest mates to their feeder via waggle dancing; $\gamma$ min$^{-1}$ is the rate bees spontaneously abandon a feeder\footnote{We have associated units of min$^{-1}$ with interaction rates. Though $\alpha_{A,B}(t)$ are in fact food qualities (See Table~\ref{table1} in Appendix), we assume the commitment term also carries units of min$^{-1}$ via a unit rescaling, which we do not include in Eq.~(\ref{2siteswarm}) to keep it from becoming too cumbersome. We make a similar assumption for the single feeder model.}; and ${\mc S}(x,y)$ is a nonlinear function describing inhibitory social interactions (e.g., stop-signaling, direct switching as described in Appendix~\ref{socialin}). Since the swarm commitment fractions are bounded within the simplex $0 \leq u_{A,B} \leq 1$ and $0 \leq u_A + u_B \leq 1$, the commitment ($\alpha_{A,B}$) and recruitment ($\beta$) provide positive feedback and the abandonment ($\gamma$) and inhibition (${\mc S}$) provide negative feedback.

Foraging efficacy is quantified by the reward rate (RR) of the swarm, assuming net nutrition is proportional to both the fraction of the swarm at a feeder $u_X$ times the current quality of that feeder minus the foraging cost $c$ (e.g., the energy required to forage and/or the predator risk), $\alpha_X(t) - c$. Integrating this product and scaling by time yields the effective RR:
\begin{align}
J(\alpha_{A,B}(t), \beta, \gamma, {\mc S}) = \frac{1}{T_f} \int_0^{T_f} \left[ u_A(t) \cdot (\alpha_A(t) - c) + u_B(t) \cdot (\alpha_B(t) - c) \right] \d t.   \label{rr2}
\end{align}
Given a food quality switching schedule $\alpha_{A,B}(t)$ and total foraging time $T_f$, swarms with more efficient foraging strategies $(\beta, \gamma, {\mc S})$ have higher RRs $J$.

Before studying how social inhibition shapes swarm foraging in two feeder environments, we analyze the single feeder model, finding that commitment and negative feedback from either abandonment or inhibition are usually sufficient for the swarm to rapidly adapt to feeder quality switches.

\subsection{Shaping swarm adaptivity and consensus for single feeders} 

\begin{figure}[t!]
\begin{center} \includegraphics[width = 1\linewidth]{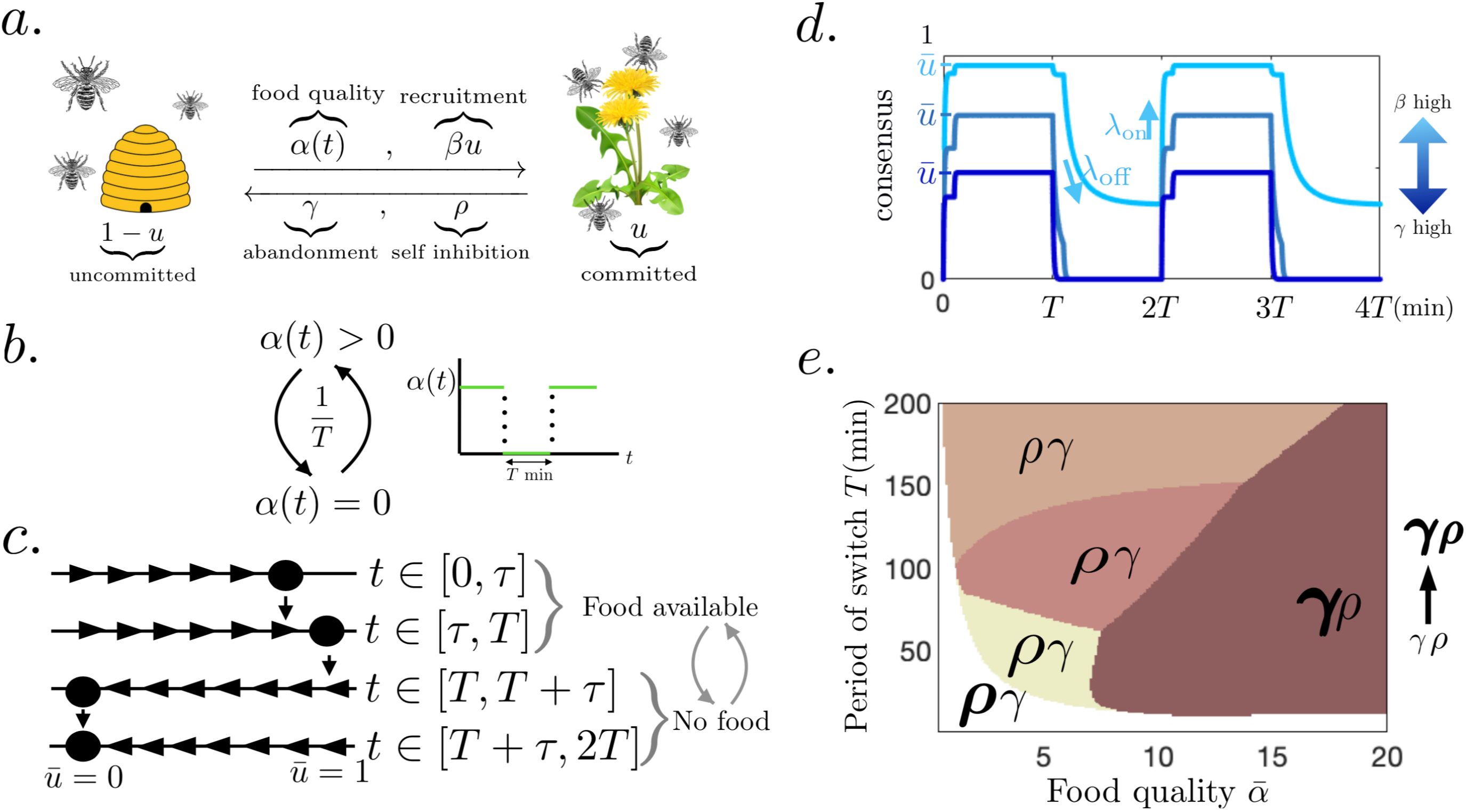} \end{center}
\caption{Swarm dynamics in the single feeding site model. ({\bf a}) Schematic of swarm foraging single site, Eq.~(\ref{singledyn}). ({\bf b}) Food availability $\alpha(t)$ switches on $\balpha$ and off $0$ at time intervals $T$ (min). ({\bf c}) Phase line plots: Equilibria of Eq.~(\ref{singledyn}) within each food quality epoch are marked as dots. Dynamic increases/decreases of the foraging fraction are indicated by right/left arrows. Bees forage when food becomes available ($\alpha \to \bar{\alpha} > 0$ for $t \in [0,T)$) and $\bar{u} > 0$ is stable and abandon the feeder once food is removed ($\alpha \to 0$ for $t \in [T, 2T)$) if recruitment is weaker than abandonment ($\beta < \gamma$). ({\bf d}) Swarm fraction foraging $u(t)$ tracks environmental changes. Higher/lower consensus $\bar{u}$ is obtained by changing the balance of recruitment $\beta$ and abandonment $\gamma$. ({\bf e}) Reward rate maximizing strategies vary with feeder quality ($\bar{\alpha}$) and switching interval ($T$). Each colored region denotes a different optimal strategy given the environment $(\bar{\alpha}, T)$. The best strategies exclude recruitment ($\beta =0$). Boldness of letters $\gamma$ and $\rho$ denote the strength of swarm behaviors that best adapt to the given environment. In rapid (short $T$) or low quality (low $\bar{\alpha}$) environments (white region), strong inhibition $\rho$ and weak abandonment $\gamma$ is best, whereas in slow or high quality environments, inhibition $\rho$ can be weak. We take $\tau = T/10$ min throughout. See Appendix~\ref{optimone} for optimization methods. }  
\label{fig2}
\end{figure}

Inhibitory social interactions in a single feeder model can only take the form of {\em self inhibition}, by which a foraging bee stops another based on a detected change in food quality (Fig.~\ref{fig2}a). Since transit from the hive to the feeder takes time, we incorporate a delay of $\tau$ minutes, so the fraction of foraging bees $u$ evolves as:
\begin{align} \label{singledyn}
\dot{u} &= (1-u) ( \alpha (t) + \beta u) - \gamma u - \rho(\balpha - \alpha(t-\tau))u^2,  
\end{align}
where $\alpha (t)$ is the food quality schedule of the feeder that switches at time intervals $T$ (minutes) between $\alpha(t) =0$ and $\alpha(t) = \bar{\alpha}$~\cite{seeley91,granovskiy12} (Fig.~\ref{fig2}b), $\beta$ min$^{-1}$ and $\gamma$ min$^{-1}$ are the recruitment and abandonment rates, and $\rho$ min$^{-1}$ is the rate of self-inhibition.

Swarm adaptivity and consensus is shaped by both individual behavior changes (commitment $\alpha (t)$ and abandonment $\gamma$) and interactions (recruitment $\beta$ and inhibition $\rho$)~\cite{seeley12}. Periodic solutions to Eq.~(\ref{singledyn}) can be found explicitly, allowing us to to compute a swarm's reward rate (RR)
(See Appendix \ref{evolnmean}).
Adaptive swarms rapidly return to the hive when no food is available and quickly populate the feeder when there is food (Fig.~\ref{fig2}c,d).
Eq.~(\ref{singledyn}) admits one stable equilibrium in each time interval: When no food is available ($\alpha(t) =0$) the nonforaging ($\bar{u}=0$) equilibrium is stable as long as recruitment is not stronger than abandonment ($\beta < \gamma$). When food becomes available ($\alpha (t) = \bar{\alpha} >0$) the stable fraction of foragers $\bar{u}$ increases with food quality (See Fig.~\ref{fig2}c and Appendix \ref{equilanaz}). This fraction $\bar{u}$ corresponds to the {\em consensus} of the swarm~\cite{conradt05}, and the rate $\lambda$ we deem the swarm's {\em adaptivity}.

\subsection{Robust foraging should adapt to the environmental conditions}

The performance of swarm interaction strategies strongly depends on the feeder quality $\bar{\alpha}$ and switching time $T$.
Swarms with stronger rates of abandonment $\gamma$ and self-inhibition $\rho$ more quickly leave the feeder once there is no food ($\alpha(t): \bar{\alpha} \mapsto 0$), but have limited consensus $\bar{u}$ when food becomes available ($\alpha(t): 0 \mapsto \balpha$). Increasing the recruitment rate $\beta$, on the other hand, boosts consensus but can slow the rate at which the swarm abandons an empty feeder (Fig.~\ref{fig2}d).

To quantify the effect of abandonment $\gamma$, recruitment $\beta$, and self-inhibition $\rho$, we compute the long term RR of the swarm, measuring the foraging yield over a single period ($2T$ minutes) once the swarm equilibrates to its periodic switching behavior (See Appendix~\ref{evolnmean}):
\begin{align} 
J(\gamma, \beta, \rho) = \frac{1}{2T} \int_0^{2T} (\alpha(t) - c) u(t) \d t.  	\label{rr1}
\end{align}
where $0<c < \bar{\alpha}$ is the cost of foraging and $\alpha(t)\in \{ 0, \bar{\alpha}\}$ is the quality of foraging site.

For each foraging site quality level, $\bar{\alpha}$, there is an optimal foraging strategy (abandonment $\gamma$, recruitment $\beta$ and stop signaling $\rho$) within our set of possible strategies (See Appendix~\ref{optimone}) that maximizes the RR $J(\gamma, \beta, \rho)$ (Fig.~\ref{fig2}e). Here, private information is sufficient for individual bees to commit to foraging (quality sensing $\alpha (t)$), and recruitment does not benefit the swarm ($\beta = 0$). Reinforcing the majority opinion via recruitment is detrimental once the environment changes, as opposed to static environments~\cite{camazine99,franks02,seeley12}. In rapid (small $T$) or low food quality ($\bar{\alpha}$ low) environments, stronger inhibition (large $\rho$) is needed to swap swarm commitment when the environment changes (white region, Fig.~\ref{fig2}e). This nonlinear mechanism increases the adaptivity of the swarm, but tempers the initial stage of consensus after the feeder is switched on (See Appendix~\ref{equilanaz} for details). On the other hand, when food is plentiful (high $\bar{\alpha}$) (brown regions, Fig.~\ref{fig2}e), inhibition should be weak (small $\rho$). In intermediate environments, the best strategies interpolate these extremes.

Linearizing solutions to the model Eq.~(\ref{singledyn}) provides us with a closer look at how swarm dynamics impact foraging yields. In sufficiently slow environments (large $T$) with small delays ($\tau \to 0$), we can linearly approximate the evolving foraging fraction (See Appendix~\ref{linearappone}):
\begin{align}
u(t) \approx \left\{ \begin{array}{cc} \bar{u}(1 - \e^{-\lambda_{\rm on} t}), & t \in [0,T), \\ \bar{u} \e^{- \lambda_{\rm off} t}, & t \in [T, 2T), \end{array} \right.   \label{uoneprox}
\end{align}
where $\bar{u}$ is the consensus foraging fraction and $\lambda_{{\rm on}/{\rm off}}$ are the rates the swarm arrives/departs the feeder once food is switched on/off. Plugging Eq.~(\ref{uoneprox}) into Eq.~(\ref{rr1}), we estimate the RR:
\begin{align}
J \approx \frac{\bar{u}}{2} \left[ (\bar{\alpha} - c) \left( 1 - \frac{1- \e^{- \lambda_{\rm on} T}}{\lambda_{\rm on} T} \right) - c \frac{1 - \e^{- \lambda_{\rm off} T}}{\lambda_{\rm off} T} \right].  \label{linJone}
\end{align}
It can be shown that $\pd_{\lambda} J > 0$ for $\lambda= \lambda_{{\rm on}/{\rm off}}$, so the RR increases with the rates at which the swarm switches behaviors. These rates increase as abandonment $\gamma$ and social inhibition $\rho$ are strengthened (Appendix~\ref{equilanaz}). Clearly, $J$ increases with $\bar{u}$ since more bees forage when food is available. Increasing abandonment $\gamma$ tends to decrease consensus, so the most robust foraging strategies cannot use abandonment that is too rapid (Appendix~\ref{equilanaz}).

We conclude that the volatility ($1/T$) and profitability ($\bar{\alpha}$) of the environment dictate the swarm interactions that yield efficient foraging strategies. One important caveat is that we bounded the interaction parameters, so swarm communication cannot be arbitrarily fast. This biological bound may be lower in practice, explaining slow adaptation of swarms to feeder changes in experiments~\cite{seeley91,granovskiy12}. Our qualitative finding, that social inhibition is more effective in slow and high quality environments, should be robust to even tighter bounds. We have also shown that when social inhibition is not present, abandonment must be increased as the speed and quality of the environment is increased (Appendix~\ref{abandon} and Fig.~\ref{fig7}). In the next section, we extend these principles to two feeder environments, particularly showing how specific forms of social inhibition shape foraging yields.

\subsection{Foraging decisions between two dynamic feeders}

For the swarm to effectively decide between two feeders, it must collectively inhibit foraging at the lower quality feeder. Our mean field model, Eq.~(\ref{2siteswarm}), generalizes house-hunting swarm models with stop-signaling~\cite{franks02,seeley12, pais13} to a foraging swarm in a dynamic environment with different forms of social inhibition~(Fig.~\ref{fig1}). How do these inhibitory interactions contribute to foraging efficacy? Honeybees can deliver inhibitory signals to nestmates foraging at potentially perilous sites~\cite{nieh93,nieh10, pastor05}, but swarm-level effects of these mechanisms are not well studied in foraging tasks in dynamic environments~\cite{tan16}. As we will show, the specific form of social inhibition can strongly determine how a swarm will adapt to change.

\begin{figure}[t!]
\begin{center}
{\includegraphics[width = 15.5cm]{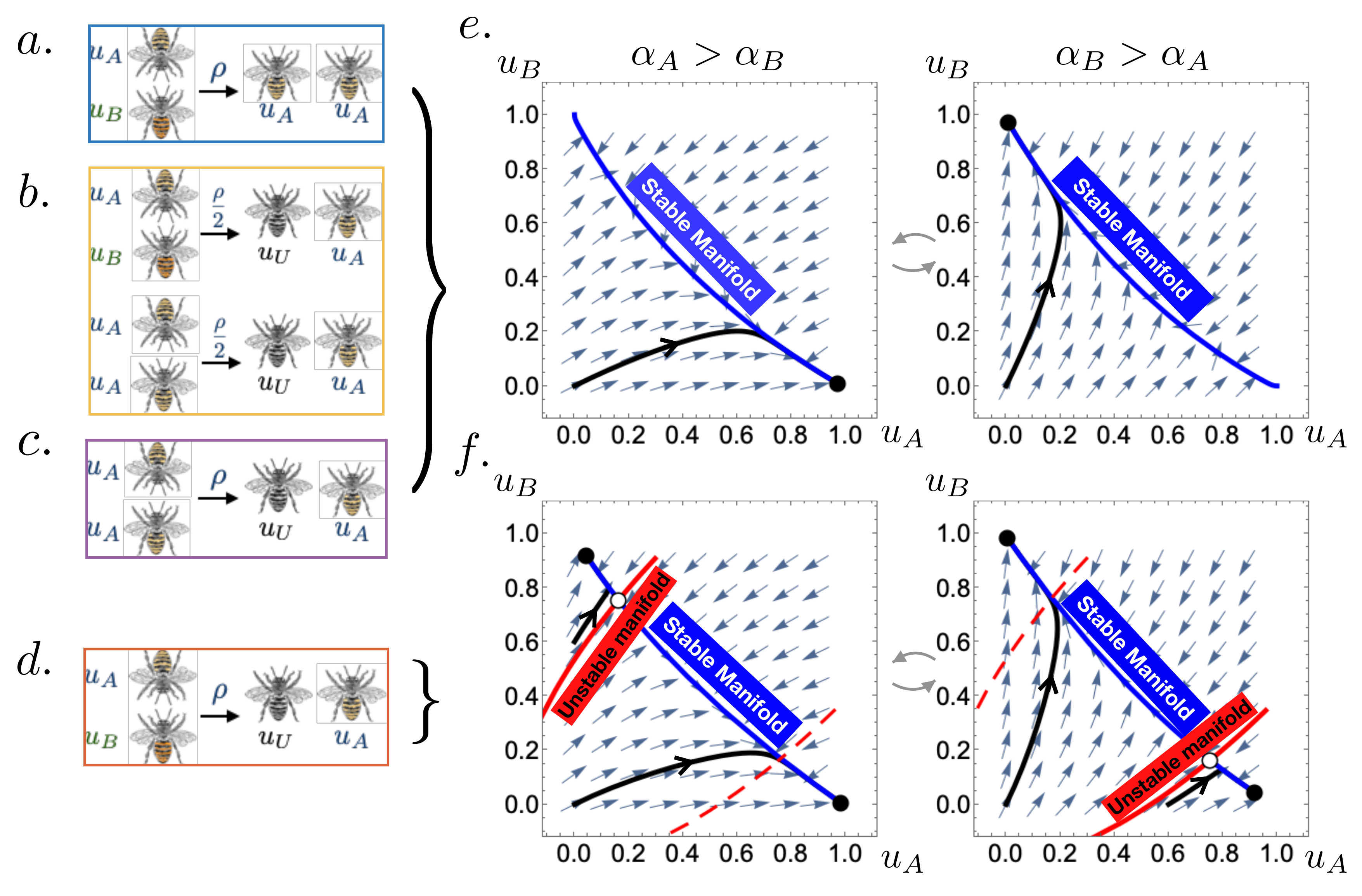}}
\caption{Social inhibition of swarm foraging two feeders (See Fig.~\ref{fig1}) and their resultant dynamical behaviors: ({\bf a}) direct switching: inhibiting bee switches opposing nestmate to their preference; ({\bf b}) indiscriminate stop signaling: inhibiting bee may either cause opposing or agreeing nestmate to become uncommitted; ({\bf c}) self-inhibition: inhibiting bee causes agreeing nestmate to become uncommitted; ({\bf d}) discriminate stop signaling: inhibiting bee causes opposing nestmate to become uncommitted. Phase portraits: ({\bf e}) Monostable behavior arises from direct switching, discriminate stop signaling, and self-inhibition so the swarm always tends to a single equilibrium foraging fraction given fixed feeder qualities. ({\bf f}) Bistable behavior that can emerge for strong discriminate stop-signaling. }
\label{fig3}
\end{center}
\end{figure}

\subsection{Forms of social inhibition} Generalizing previous models~\cite{britton02,seeley12}, we consider four forms of social inhibition (all parameterized by $\rho$ as before): (a) direct switching: bees foraging at the superior feeder directly switch the preference of opposing foragers to the better feeder; (b) indiscriminate stop-signaling: when two foraging bees meet, one will stop foraging; (c) self-inhibition: when two bee foraging at the same feeder meet, one will stop foraging; and (d) discriminate stop-signaling: when bees foraging different feeders meet, one stops foraging. These interactions are visualized in Fig.~\ref{fig3}a,b,c,d and their evolution equations are given in Appendix~\ref{socialin} (See also \cite{seeley12}~supplement).

We can divide these forms of social inhibition into two classes, based on the swarm dynamics they produce: monostable or bistable consensus behaviors. The first three forms of social inhibition yield swarms with monostable consensus behaviors (See Appendix~\ref{twolinstab}), tending to a single stable foraging fraction when the feeder qualities are fixed (Fig.~\ref{fig3}e). The swarm will thus mostly forage at the higher yielding feeder. On the other hand, strong discriminant stop-signaling can produce swarms with bistable consensus behaviors (Fig.~\ref{fig3}f). As a result, the swarm can remain stuck at an unfavorable feeder, after the feeder qualities are switched. This is similar to ``winner-take-all" regimes in mutually inhibitory neural networks~\cite{wong06,marshall09}. Inhibition from bees holding the swarm's dominant preference is too strong for bees with the opposing preference to overcome, even with new evidence from the changed environment.

\subsection{Direct switching leads to most robust foraging}

\begin{figure}[t!]
\begin{center}
{\includegraphics[width = 13cm]{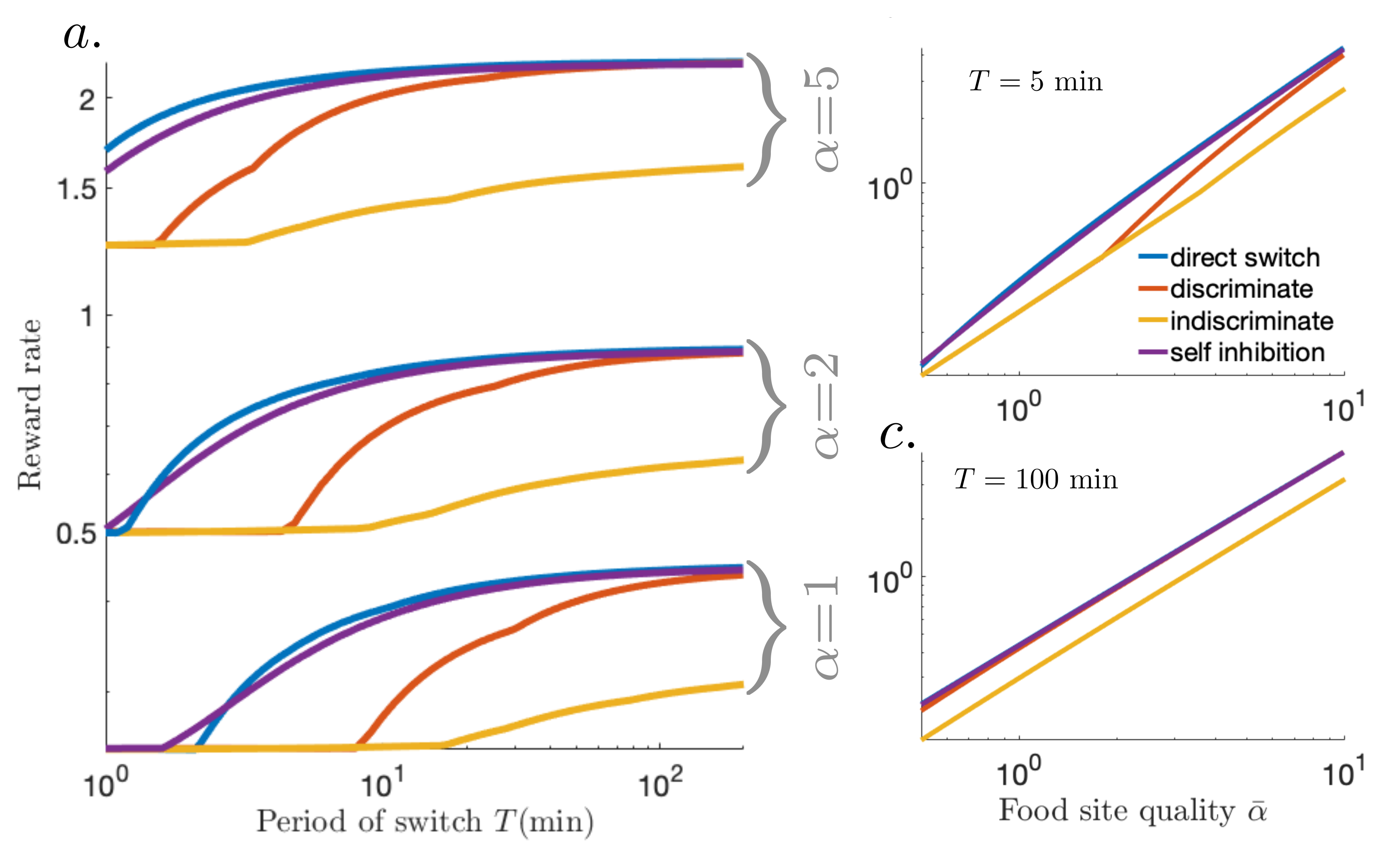}}
\caption{Optimal reward rates across forms of social inhibition. ({\bf a}) Reward rate (RR) increases as the interval between feeder quality switches $T$ increases for all social inhibition strategies. Across most parameter sets, direct switching is the most robust strategy, yielding the highest RRs. In rapid environments, self-inhibition can be slightly better.  ({\bf b}) For $T=5$~min fixed and maximal food quality $\bar{\alpha}$ varied, RRs for direct switching and self inhibition are separated from discriminate and indiscriminate stop-signaling at lower food quality levels $\bar{\alpha}$. ({\bf c})  For $T= 100$~min, direct switching, discriminate stop-signaling, and self-inhibition yield similar RRs, whereas indiscriminate stop-signaling is notably worse. These curves are fit near perfectly by a linear function ($R^2 = 1.0000$). } \label{fig4}
\end{center}
\end{figure}

To determine the most robust forms of social inhibition for foraging in dynamic environments, we studied how the rate of reward, Eq.~(\ref{rr2}), depended on the foraging strategy used. Environments are parameterized by the time between switches $T$ (min), the better feeder quality $\bar{\alpha}$ and the lower feeder quality $\bar{\alpha}/2$, which periodically switch between feeders $A$ and $B$. As in the single feeder case, we tune interactions of each strategy (Fig.~\ref{fig3}a,b,c,d) to maximize reward rate (RR) over a discrete set of strategies (See Appendix~\ref{optimtwo} for details). Comparing each social inhibition strategy type's RR in different environments (Fig.~\ref{fig4}a), we find direct switching generally yields higher RRs than other strategies. Deviations between the effectiveness of different strategies are most pronounced at intermediate environmental timescales $T$. As expected, RRs increase with the maximal food site quality $\bar{\alpha}$ (Fig.~\ref{fig4}b,c).

Direct switching is likely a superior strategy because it allows for continually foraging (Fig.~\ref{fig3}a), rather than other strategies' interruption by an uncommitted stage (Fig.~\ref{fig3}b,c,d), which must rely on recruitment $\beta$ to restart foraging. To study how interactions should be balance to yield effective foraging, we examined how to optimally tune $(\beta, \gamma, \rho)$ across environments in the direct switching model (Fig.~\ref{fig5}). Analysis of other models are show in Figs \ref{fig8} and \ref{fig9} of Appendix~\ref{foragtune}.

\begin{figure}[t!]
\begin{center}
{\includegraphics[width = 1\linewidth]{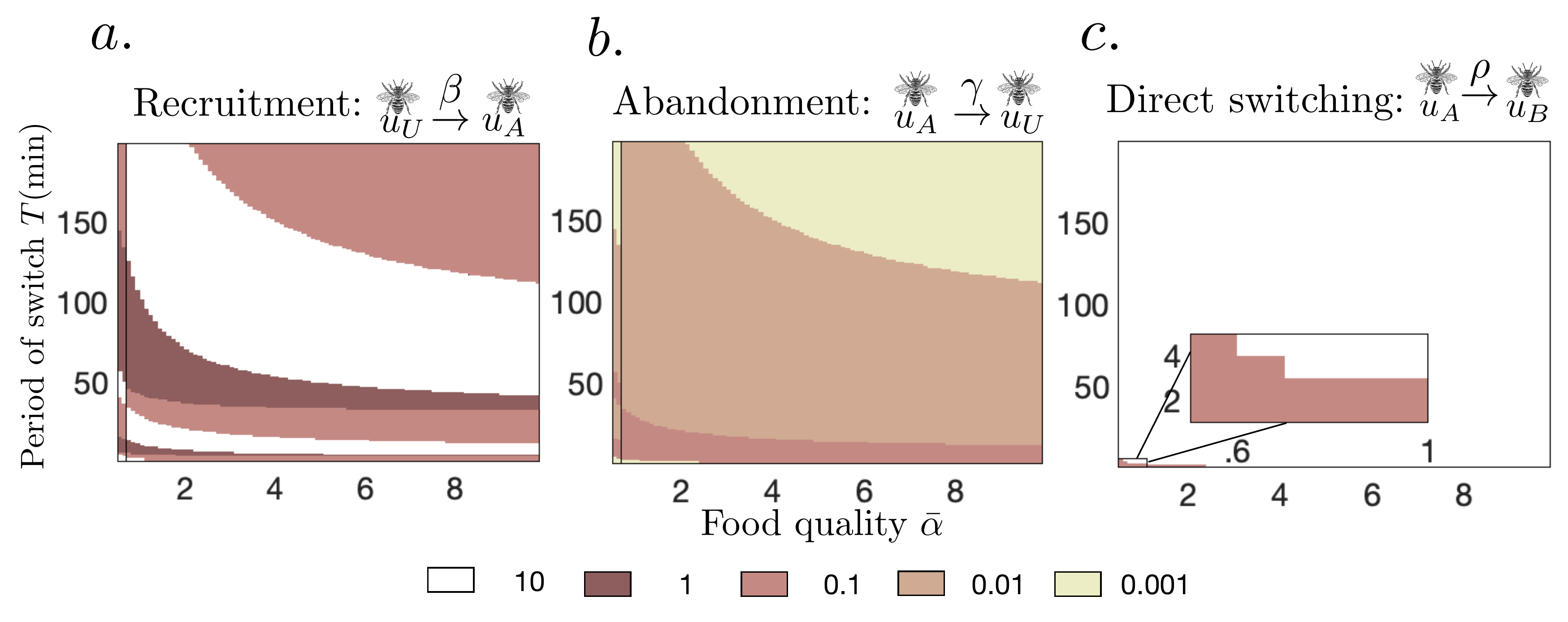}}
\caption{Tuning ({\bf a}) recruitment $\beta$; ({\bf b}) abandonment $\gamma$; and ({\bf c}) social inhibition $\rho$ to maximize RR in the direct switching model (Fig.~\ref{fig3}a). See Appendix~\ref{optimtwo} for methods. ({\bf a}) The best tunings of recruitment vary considerably for rapid (low $T$) and low quality $\bar{\alpha}$ environments, but recruitment appears to be less essential for slow (high $T$) and high quality $\bar{\alpha}$ environments. ({\bf b}) The rate of abandonment that best suits the environment decreases as the environment becomes slower and higher quality. ({\bf c}) Generally, the direct switching rate should be made as strong as possible, except for very fast, low quality environments.}
\label{fig5}
\end{center}
\end{figure}

As in the single feeder environments, we see a delineation between strategies optimized to slow/high quality environments as opposed to rapid/low quality environments. Weak recruitment $\beta$ (Fig.~\ref{fig5}a) and abandonment $\gamma$ (Fig.~\ref{fig5}b), and strong direct switching (Fig.~\ref{fig5}c) yield the highest RRs in slow (large $T$) and high quality (large $\bar{\alpha}$) environments. Recruitment $\beta$ may be inessential since the food quality signals $\bar{\alpha}$ and $\bar{\alpha}/2$ are significantly different. Also, direct switching $\rho$ provides strong adaptation to change. In fact, for virtually all environments, we found it was best to take $\rho$ as strong as possible. The strategy changes significantly when the environment is fast (small $T$) and low quality (small $\bar{\alpha}$), in which case abandonment $\gamma$ should be strong, and in extreme cases direct switching $\rho$ can be made weak (Fig.~\ref{fig5}b,c). Changes in the optimal recruitment strength are less systematic, however, and there are stratified regions in which the best $\beta$ can change significantly for small shifts in environmental parameters. Overall, a mixture of abandonment and direct switching is more effective in more difficult environments (lower $T$ and $\bar{\alpha}$).


Direct switching does underperform self-inhibition in rapid environments (Fig.~\ref{fig4}a), since the swarm is more efficient by keeping some bees uncommitted, and not risking the cost of foraging the lower yielding feeder. Strong self-inhibition $\rho$ keeps more bees from foraging. Overall, both direct switching and self-inhibition can perform similarly, as recruitment interactions can be strengthened in self-inhibiting swarms, so more bees return to foraging after such inhibitory encounters (Fig.~\ref{fig4}). This balances adaptivity, so the swarm's preferences change with the environment, and consensus, so the swarm mostly builds up to forage at the better feeder given sufficient time. We now study this balance in each model using linearization techniques. Overall, these measures can account for discrepancies between the RR yields of swarms using different social inhibition strategies.

\subsection{Linearization reveals strategy adaptivity and consensus}
Each swarm interaction mechanism differentially shapes both the fraction of bees that forage at the better site in the long time limit (consensus $\bar{u}$) and the rate at which this bound is approached (adaptivity $\lambda$). Focusing specifically on these measures, we demonstrate both how they shape foraging efficiency and how they distinguish each social inhibition strategy.

\begin{figure}[t]
\begin{center} 
{\includegraphics[width = 1\linewidth]{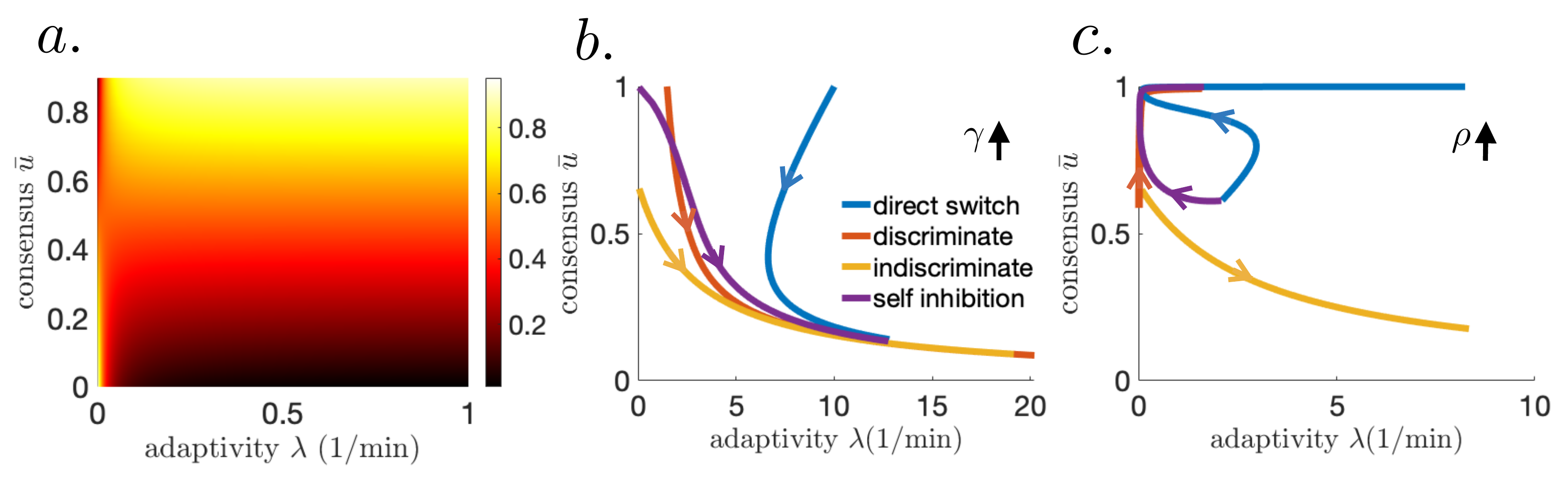}}
\caption{ ({\bf a}) Foraging yield varies with consensus ($\bar{u}$) and adaptivity ($\lambda$) when  $T=100$~min. Tradeoff between consensus and adaptivity in the linearized model for ({\bf b})  abandonment, $\gamma$ between $[0,20]$ $\text{min}^{-1}$ ({\bf c})  social inhibition, $\rho$ between $[0,20]$ $\text{min}^{-1}$. Switching time $T=100$ min, food quality $\bar{\alpha} = 2$, and other parameters are fixed at their optimal level. }		\label{fig6}
\end{center}
\end{figure}

We leverage our approach developed for the single feeder model, and consider linear approximations of Eq.~(\ref{2siteswarm}) in the limit of long switching times $T$ (See Appendix~\ref{linearapptwo} and Fig.~\ref{fig10} in Appendix~\ref{linac}). In the specific case $c : = \bar{\alpha}/2$, we can approximate the RR solely in terms of the consensus $\bar{u}$ (long term fraction of the swarm at the better feeder) and adaptivity $\lambda$ (rate this fraction is approached):
\begin{align}
J \approx \frac{\bar{\alpha}}{2} \left( \bar{u} + (1 - 2 \bar{u}) \frac{1 - \e^{- \lambda T}}{\lambda T} \right).	\label{twofeedj}
\end{align}
The RR $J$ increases with consensus $\bar{u}$ and adaptivity $\lambda$ (Fig.~\ref{fig6}a). Efficient swarms rapidly recruit a high fraction of the swarm to the better feeder. Consensus and adaptivity are approximated in each model using linear stability (Appendix~\ref{twolinstab}). The impact of varying abandonment $\gamma$ and social inhibition $\rho$ on $\bar{u}$ and $\lambda$ is consistent with our optimality analysis of the full nonlinear model (Fig.~\ref{fig6}b,c): Social inhibition generates more robust swarm switching between feeders than abandonment. While strengthening abandonment $\gamma$ can increase $\lambda$, it decreases consensus $\bar{u}$ since it causes bees to become uncommitted (Fig.~\ref{fig6}b). Such consensus-adaptivity trade-offs do not occur in most models, as social inhibition $\rho$ is strengthened (See also Figs \ref{fig11} and \ref{fig12} in Appendix~\ref{adconmod}). Only indiscriminate stop-signaling exhibits this behavior (Fig.~\ref{fig6}c), but the other three models (direct switching, discriminate stop signaling and self inhibition) do not. Rather, consensus $\bar{u}$ increases with social inhibition, while adaptivity can vary nonmonotonically (direct switching) or even decrease (self-inhibition). Overall, direct switching swarms attain the highest levels of consensus and adaptivity, consistent with our finding that it is the most robust model (Fig.~\ref{fig4}).


Direct switching sustains both high levels of consensus $\bar{u}$ and adaptivity $\lambda$ (Fig.~\ref{fig6}c; See also Figs \ref{fig11} and \ref{fig12} in Appendix~\ref{adconmod}). The resulting swarms quickly discard their prior beliefs about the highest yielding feeder, exhibiting leaky evidence accumulation~\cite{glaze15}. On the other hand, strong abandonment $\gamma$ (Fig.~\ref{fig6}b) or indiscriminate stop signaling (Fig.~\ref{fig6}c) increase adaptivity but limit consensus $\bar{u}$ at the better feeder. Strengthening recruitment $\beta$ leads to stronger consensus $\bar{u}$ at the expense of adaptivity $\lambda$ to environmental changes. Swarms likely use a combination of social inhibition and abandonment mechanisms~\cite{jack15}, so such consensus-adaptivity trade-offs are important to manage in dynamic environments.

\section{Discussion}

Foraging animals constantly encounter temporal and spatial changes in their food supply~\cite{owen10}. The success of a foraging animal groups thus depends on how efficiently they communicate and act upon environmental changes~\cite{grueter12}. Our swarm model analysis pinpoints specific social inhibition mechanisms that facilitate adaptation to changes in food availability and consolidate consensus at better feeding sites. If bees interact by direct switching, they can immediately update their foraging preference without requiring recruitment, keeping foragers active following environmental changes. Recruitment is less important to the foraging success of a swarm in dynamic conditions; bees can initiate commitment via their own scouting behavior. Individuals should balance their social and private information in an environment-dependent way to decide and forage most efficiently~\cite{czaczkes11,torney15}. 

Efficient group decision making combines individual private evidence accumulation and information sharing across the cohort~\cite{conradt08}. However, in groups where social influence is strong, opinions generated from weak and potentially misleading private evidence can cascade through the collective, resulting in rapid but lower value decisions~\cite{dall05,sumpter08,pais13}. Our analysis makes these notions quantitatively concrete by associating the accuracy of the swarm decisions with the consensus fraction at the better feeder and the speed of decisions with the adaptivity or rate the swarm approaches steady state consensus (Fig.~\ref{fig6}). The best foraging strategies balance these swarm level measures of decision efficiency. Social insects do appear to balance the speed and accuracy of decision to increase their rate of food intake~\cite{chittka03,burns08}, and collective tuning is likely influenced by individuals varying their response to social information.


We find that social recruitment can speed along initial foraging decisions, but it can limit adaptation to change. This is consistent with experimental studies that show a reduction in positive feedback can help collectives steer away from lower value decisions.
For example, challenging environmental conditions (e.g., volatile and low food quality) are best managed by honeybee swarms whose individuals do not wait for recruitment but rely on their own individual scouting~\cite{price19}. Ants encountering crowded environments tend to deposit less pheromone to keep their nestmates from less efficient foraging paths~\cite{czaczkes13}. These experimental findings suggest social insects adapt to changing environmental conditions by limiting communication that promotes positive feedback~\cite{grueter14}. Foragers must then be proactive in dynamic environments, since they cannot afford to wait for new social information~\cite{dechaume05}.
Thus, the advantages of social learning depend strongly on environmental conditions~\cite{laland04}.

In concert with a reduction in recruitment, we predict that honeybee swarms foraging in volatile environments will benefit from strengthening inhibitory mechanisms at the individual and group level. Bees enacting social inhibition dissuade their nestmates from foraging at opposing feeders. We found the most efficient form of social inhibition is direct switching whereby bees flip the opinion of committed bees to their own opinion. So do honeybees utilize this mechanism in dynamic environments?
Observations of swarms making nest site decisions show scouts directly switching their dance allegiance~\cite{gould94,britton02}, but these events seem to be relatively rare in the static environments of typical nest site selection experiments~\cite{camazine99}. Other forms of social inhibitory signals, especially stop-signaling, appear to be used to promote consensus in nest decisions~\cite{seeley12,pais13} and escaping predators while foraging \cite{nieh93,nieh10}. Thus, the role and prevalence of social inhibition as a means for foraging adaptively in dynamic environments warrants further investigation.

Our simple parameterized model, developed from previously validated house-hunting models~\cite{franks02,marshall09,seeley12}, is amenable to analysis and could be validated with time series measurements from dynamic foraging experiments. Past experimental work focused on shorter time windows in which only a few switches in feeder quality occurred~\cite{seeley91,granovskiy12}, which may account for the relatively slow adaptation of the swarms to environmental changes. We predict swarms will slowly tune their social learning strategies to suit the volatility of the environment, but this could require several switch observations. Foraging tasks conducted within a laboratory could be controlled to track bee interactions over long time periods using newly developed automated monitoring techniques~\cite{gernat18}. Our study also identifies key regions in parameter space in which different foraging strategies diverge in their performance, suggesting that placing swarms in rapid environments with relatively low food supplies will help distinguish which social communication mechanisms are being used.

Collective decision strategies and outcomes can depend on group size~\cite{hill82,krause10}, though decision accuracy does not necessarily increase with group size~\cite{kao14}. We approximated swarm dynamics using a population level model, which is the deterministic mean field limit of a stochastic individual interaction based model~\cite{seeley12}. Finite group size considerations would result in a models which behaves stochastically, in which the same conditions can generate different swarm dynamics~\cite{pais13}. The qualitative predictions of our mean field generally did not change dramatically when considering the stochastic finite size model (See Appendix~\ref{stochsys} and Fig.~\ref{fig13}). However discriminate stop-signaling swarms exhibit bistable decision dynamics (Fig.~\ref{fig3}d,f), so we expect in the stochasticity in the finite sized model would allow swarms to break free from less profitable feeders, similar to noise-driven escapes of a particle in double potential well models~\cite{gammaitoni98}. Fluctuation-induced switching may provide an additional mechanism for flexible foraging~\cite{dussutour09,biancalani14}, and would be an interesting extension of our present modeling work.

\ \\
\vspace{-3mm}



\noindent
{\bf Data Accessibility:} Code for producing figures is available at \href{https://github.com/sbidari/dynamicbees}{https://github.com/sbidari/dynamicbees} \\[1ex]
\noindent
{\bf Authors' contributions:} Formulated scientific question, modeling approaches, developed models: SB, OP, ZPK; implemented mathematical analysis and computer simulations: SB; wrote the article: SB, OP, ZPK. All authors were involved in discussions on different aspects of the study. \\[1ex]
\noindent
{\bf Competing interests:} We declare we have no competing interests. \\[1ex]
\noindent
{\bf Funding:} SB and ZPK were supported by an NSF grant (DMS-1615737). SB was also supported by a Dissertation Fellowship from the American Association of University Women. ZPK was also supported by NSF/NIH CRCNS grant(R01MH115557). \\[1ex]
\noindent
{\bf Acknowledgements:} We thank Tahra Eissa for feedback on a draft of this manuscript.

\bibliographystyle{unsrt}

\begin{thebibliography}{10}

\bibitem{sumpter03}
D~Sumpter and Stephen Pratt.
\newblock A modelling framework for understanding social insect foraging.
\newblock {\em Behavioral Ecology and Sociobiology}, 53(3):131--144, 2003.

\bibitem{visscher07}
P~Kirk Visscher.
\newblock Group decision making in nest-site selection among social insects.
\newblock {\em Annual Review of Entomology}, 52(1):255--275, 2007.

\bibitem{holldobler08}
Bert Holldobler and E~O Wilson.
\newblock {\em The Superorganism: The Beauty, Elegance, and Strangeness of
  Insect Societies}.
\newblock W. W. Norton \& Company; 1 edition, 2009.

\bibitem{ward08}
Ashley~JW Ward, David~JT Sumpter, Iain~D Couzin, Paul~JB Hart, and Jens Krause.
\newblock Quorum decision-making facilitates information transfer in fish
  shoals.
\newblock {\em Proceedings of the National Academy of Sciences},
  105(19):6948--6953, 2008.

\bibitem{ame06}
Jean-Marc Am{\'e}, Jos{\'e} Halloy, Colette Rivault, Claire Detrain, and
  Jean~Louis Deneubourg.
\newblock Collegial decision making based on social amplification leads to
  optimal group formation.
\newblock {\em Proceedings of the National Academy of Sciences},
  103(15):5835--5840, 2006.

\bibitem{seeley10}
Thomas~D Seeley.
\newblock {\em Honeybee Democracy}.
\newblock Princeton Univ. Press, Princeton, NJ, 2010.

\bibitem{seeley00}
Thomas~D Seeley, Alexander~S Mikheyev, and Gary~J Pagano.
\newblock Dancing bees tune both duration and rate of waggle-run production in
  relation to nectar-source profitability.
\newblock {\em Journal of Comparative Physiology A}, 186(9):813--819, 2000.

\bibitem{nieh10}
James~C Nieh.
\newblock A negative feedback signal that is triggered by peril curbs honey bee
  recruitment.
\newblock {\em Current Biology}, 20(4):310--315, 2010.

\bibitem{seeley12}
TD~Seeley, PK~Visscher, T~Schlegel, PM~Hogan, NR~Franks, and JA~Marshall.
\newblock Stop signals provide cross inhibition in collective decision-making
  by honeybee swarms.
\newblock {\em Science (New York, NY)}, 335(6064):108, 2012.

\bibitem{pais13}
Darren Pais, Patrick~M Hogan, Thomas Schlegel, Nigel~R Franks, Naomi~E Leonard,
  and James~AR Marshall.
\newblock A mechanism for value-sensitive decision-making.
\newblock {\em PloS one}, 8(9):e73216, 2013.

\bibitem{cinquin02}
Olivier Cinquin and Jacques Demongeot.
\newblock Positive and negative feedback: striking a balance between necessary
  antagonists.
\newblock {\em Journal of theoretical biology}, 216(2):229--241, 2002.

\bibitem{garnier07}
Simon Garnier, Jacques Gautrais, and Guy Theraulaz.
\newblock The biological principles of swarm intelligence.
\newblock {\em Swarm intelligence}, 1(1):3--31, 2007.

\bibitem{real88}
Leslie Real and Beverly~J Rathcke.
\newblock Patterns of individual variability in floral resources.
\newblock {\em Ecology}, 69(3):728--735, 1988.

\bibitem{fewell99}
Jennifer~H Fewell and Susan~M Bertram.
\newblock Division of labor in a dynamic environment: response by honeybees
  (apis mellifera) to graded changes in colony pollen stores.
\newblock {\em Behavioral ecology and sociobiology}, 46(3):171--179, 1999.

\bibitem{seeley91}
Thomas~D Seeley, Scott Camazine, and James Sneyd.
\newblock Collective decision-making in honey bees: how colonies choose among
  nectar sources.
\newblock {\em Behavioral Ecology and Sociobiology}, 28(4):277--290, 1991.

\bibitem{benshahar02}
Y~Ben-Shahar, A~Robichon, MB~Sokolowski, and GE~Robinson.
\newblock Influence of gene action across different time scales on behavior.
\newblock {\em Science}, 296(5568):741--744, 2002.

\bibitem{tenczar14}
Paul Tenczar, Claudia~C Lutz, Vikyath~D Rao, Nigel Goldenfeld, and Gene~E
  Robinson.
\newblock Automated monitoring reveals extreme interindividual variation and
  plasticity in honeybee foraging activity levels.
\newblock {\em Animal Behaviour}, 95:41--48, 2014.

\bibitem{dornhaus04}
Anna Dornhaus and Lars Chittka.
\newblock Why do honey bees dance?
\newblock {\em Behavioral Ecology and Sociobiology}, 55(4):395--401, 2004.

\bibitem{granovskiy12}
Boris Granovskiy, Tanya Latty, Michael Duncan, David~JT Sumpter, and Madeleine
  Beekman.
\newblock How dancing honey bees keep track of changes: the role of inspector
  bees.
\newblock {\em Behavioral Ecology}, 23(3):588--596, 2012.

\bibitem{kietzman15}
Parry~M Kietzman and P~Kirk Visscher.
\newblock The anti-waggle dance: use of the stop signal as negative feedback.
\newblock {\em Frontiers in Ecology and Evolution}, 3:14, 2015.

\bibitem{gray18}
Rebecca Gray, Alessio Franci, Vaibhav Srivastava, and Naomi~Ehrich Leonard.
\newblock Multiagent decision-making dynamics inspired by honeybees.
\newblock {\em IEEE Transactions on Control of Network Systems}, 5(2):793--806,
  2018.

\bibitem{mcnamara06}
John~M McNamara, Richard~F Green, and Ola Olsson.
\newblock Bayes' theorem and its applications in animal behaviour.
\newblock {\em Oikos}, 112(2):243--251, 2006.

\bibitem{glaze15}
Christopher~M Glaze, Joseph~W Kable, and Joshua~I Gold.
\newblock Normative evidence accumulation in unpredictable environments.
\newblock {\em Elife}, 4:e08825, 2015.

\bibitem{price19}
RI'Anson Price, N~Dulex, N~Vial, C~Vincent, and C~Gr{\"u}ter.
\newblock Honeybees forage more successfully without the ``dance language'' in
  challenging environments.
\newblock {\em Science advances}, 5(2):eaat0450, 2019.

\bibitem{britton02}
NF~Britton, NR~Franks, SC~Pratt, and TD~Seeley.
\newblock Deciding on a new home: how do honeybees agree?
\newblock {\em Proceedings of the Royal Society of London. Series B: Biological
  Sciences}, 269(1498):1383--1388, 2002.

\bibitem{marshall09}
James~AR Marshall, Rafal Bogacz, Anna Dornhaus, Robert Planqu{\'e}, Tim Kovacs,
  and Nigel~R Franks.
\newblock On optimal decision-making in brains and social insect colonies.
\newblock {\em Journal of the Royal Society Interface}, 6(40):1065--1074, 2009.

\bibitem{vankampen92}
Nicolaas~Godfried Van~Kampen.
\newblock {\em Stochastic processes in physics and chemistry}.
\newblock Elsevier, 1992.

\bibitem{conradt05}
Larissa Conradt and Timothy~J Roper.
\newblock Consensus decision making in animals.
\newblock {\em Trends in ecology \& evolution}, 20(8):449--456, 2005.

\bibitem{camazine99}
Scott Camazine, PK~Visscher, Jennifer Finley, and RS~Vetter.
\newblock House-hunting by honey bee swarms: collective decisions and
  individual behaviors.
\newblock {\em Insectes Sociaux}, 46(4):348--360, 1999.

\bibitem{franks02}
Nigel~R Franks, Stephen~C Pratt, Eamonn~B Mallon, Nicholas~F Britton, and
  David~JT Sumpter.
\newblock Information flow, opinion polling and collective intelligence in
  house--hunting social insects.
\newblock {\em Philosophical Transactions of the Royal Society of London.
  Series B: Biological Sciences}, 357(1427):1567--1583, 2002.

\bibitem{nieh93}
James~C Nieh.
\newblock The stop signal of honey bees: reconsidering its message.
\newblock {\em Behavioral Ecology and Sociobiology}, 33(1):51--56, 1993.

\bibitem{pastor05}
Kristen~A Pastor and Thomas~D Seeley.
\newblock The brief piping signal of the honey bee: begging call or stop
  signal?
\newblock {\em Ethology}, 111(8):775--784, 2005.

\bibitem{tan16}
Ken Tan, Shihao Dong, Xinyu Li, Xiwen Liu, Chao Wang, Jianjun Li, and James~C
  Nieh.
\newblock Honey bee inhibitory signaling is tuned to threat severity and can
  act as a colony alarm signal.
\newblock {\em PLoS biology}, 14(3):e1002423, 2016.

\bibitem{wong06}
Kong-Fatt Wong and Xiao-Jing Wang.
\newblock A recurrent network mechanism of time integration in perceptual
  decisions.
\newblock {\em Journal of Neuroscience}, 26(4):1314--1328, 2006.

\bibitem{jack15}
Ralph~T Jack-McCollough and James~C Nieh.
\newblock Honeybees tune excitatory and inhibitory recruitment signalling to
  resource value and predation risk.
\newblock {\em Animal behaviour}, 110:9--17, 2015.

\bibitem{owen10}
N~Owen-Smith, JM~Fryxell, and EH~Merrill.
\newblock Foraging theory upscaled: the behavioural ecology of herbivore
  movement.
\newblock {\em Philosophical Transactions of the Royal Society B: Biological
  Sciences}, 365(1550):2267--2278, 2010.

\bibitem{grueter12}
Christoph Grueter, Roger Schuerch, Tomer~J Czaczkes, Keeley Taylor, Thomas
  Durance, Sam~M Jones, and Francis~LW Ratnieks.
\newblock Negative feedback enables fast and flexible collective
  decision-making in ants.
\newblock {\em PLoS One}, 7(9):e44501, 2012.

\bibitem{czaczkes11}
Tomer~J Czaczkes, Christoph Gr{\"u}ter, Sam~M Jones, and Francis~LW Ratnieks.
\newblock Synergy between social and private information increases foraging
  efficiency in ants.
\newblock {\em Biology letters}, 7(4):521--524, 2011.

\bibitem{torney15}
Colin~J Torney, Tommaso Lorenzi, Iain~D Couzin, and Simon~A Levin.
\newblock Social information use and the evolution of unresponsiveness in
  collective systems.
\newblock {\em Journal of the Royal Society Interface}, 12(103):20140893, 2015.

\bibitem{conradt08}
Larissa Conradt and Christian List.
\newblock Group decisions in humans and animals: a survey.
\newblock {\em Philosophical Transactions of the Royal Society B: Biological
  Sciences}, 364(1518):719--742, 2008.

\bibitem{dall05}
Sasha~RX Dall, Luc-Alain Giraldeau, Ola Olsson, John~M McNamara, and David~W
  Stephens.
\newblock Information and its use by animals in evolutionary ecology.
\newblock {\em Trends in ecology \& evolution}, 20(4):187--193, 2005.

\bibitem{sumpter08}
David~JT Sumpter, Jens Krause, Richard James, Iain~D Couzin, and Ashley~JW
  Ward.
\newblock Consensus decision making by fish.
\newblock {\em Current Biology}, 18(22):1773--1777, 2008.

\bibitem{chittka03}
Lars Chittka, Adrian~G Dyer, Fiola Bock, and Anna Dornhaus.
\newblock Psychophysics: bees trade off foraging speed for accuracy.
\newblock {\em Nature}, 424(6947):388, 2003.

\bibitem{burns08}
James~G Burns and Adrian~G Dyer.
\newblock Diversity of speed-accuracy strategies benefits social insects.
\newblock {\em Current biology}, 18(20):R953--R954, 2008.

\bibitem{czaczkes13}
Tomer~J Czaczkes, Christoph Gr{\"u}ter, and Francis~LW Ratnieks.
\newblock Negative feedback in ants: crowding results in less trail pheromone
  deposition.
\newblock {\em Journal of the Royal Society Interface}, 10(81):20121009, 2013.

\bibitem{grueter14}
Christoph Grueter and Ellouise Leadbeater.
\newblock Insights from insects about adaptive social information use.
\newblock {\em Trends in Ecology \& Evolution}, 29(3):177--184, 2014.

\bibitem{dechaume05}
Fran{\c{c}}ois-Xavier Dechaume-Moncharmont, Anna Dornhaus, Alasdair~I Houston,
  John~M McNamara, Edmund~J Collins, and Nigel~R Franks.
\newblock The hidden cost of information in collective foraging.
\newblock {\em Proceedings of the Royal Society B: Biological Sciences},
  272(1573):1689--1695, 2005.

\bibitem{laland04}
Kevin~N Laland.
\newblock Social learning strategies.
\newblock {\em Animal Learning \& Behavior}, 32(1):4--14, 2004.

\bibitem{gould94}
James~L Gould and Carol~Grant Gould.
\newblock {\em The animal mind}.
\newblock WH Freeman, New York, 1994.

\bibitem{gernat18}
Tim Gernat, Vikyath~D Rao, Martin Middendorf, Harry Dankowicz, Nigel
  Goldenfeld, and Gene~E Robinson.
\newblock Automated monitoring of behavior reveals bursty interaction patterns
  and rapid spreading dynamics in honeybee social networks.
\newblock {\em Proceedings of the National Academy of Sciences},
  115(7):1433--1438, 2018.

\bibitem{hill82}
Gayle~W Hill.
\newblock Group versus individual performance: Are n+ 1 heads better than one?
\newblock {\em Psychological bulletin}, 91(3):517, 1982.

\bibitem{krause10}
Jens Krause, Graeme~D Ruxton, and Stefan Krause.
\newblock Swarm intelligence in animals and humans.
\newblock {\em Trends in ecology \& evolution}, 25(1):28--34, 2010.

\bibitem{kao14}
Albert~B Kao and Iain~D Couzin.
\newblock Decision accuracy in complex environments is often maximized by small
  group sizes.
\newblock {\em Proceedings of the Royal Society B: Biological Sciences},
  281(1784):20133305, 2014.

\bibitem{gammaitoni98}
Luca Gammaitoni, Peter H{\"a}nggi, Peter Jung, and Fabio Marchesoni.
\newblock Stochastic resonance.
\newblock {\em Reviews of modern physics}, 70(1):223, 1998.

\bibitem{dussutour09}
Audrey Dussutour, Madeleine Beekman, Stamatios~C Nicolis, and Bernd Meyer.
\newblock Noise improves collective decision-making by ants in dynamic
  environments.
\newblock {\em Proceedings of the Royal Society of London B: Biological
  Sciences}, 276(1677):4353--4361, 2009.

\bibitem{biancalani14}
Tommaso Biancalani, Louise Dyson, and Alan~J McKane.
\newblock Noise-induced bistable states and their mean switching time in
  foraging colonies.
\newblock {\em Physical review letters}, 112(3):038101, 2014.

\bibitem{bernardo08}
Mario Bernardo, Chris Budd, Alan~Richard Champneys, and Piotr Kowalczyk.
\newblock {\em Piecewise-smooth dynamical systems: theory and applications},
  volume 163.
\newblock Springer Science \& Business Media, 2008.

\bibitem{strogatz18}
Steven~H Strogatz.
\newblock {\em Nonlinear Dynamics and Chaos with Student Solutions Manual: With
  Applications to Physics, Biology, Chemistry, and Engineering}.
\newblock CRC Press, 2018.

\bibitem{gillespie77}
Daniel~T Gillespie.
\newblock Exact stochastic simulation of coupled chemical reactions.
\newblock {\em The journal of physical chemistry}, 81(25):2340--2361, 1977.

\end{thebibliography}

\ \\[5ex]
\appendix
\noindent
{\bf \LARGE Appendix}
\vspace{-4mm}
\section{Swarm foraging dynamics for a single switching feeder}
Consider model Eq.~(\ref{singledyn}) for which the food quality $\alpha (t)$ switches between two values $\alpha (t) = \bar{\alpha}$ and 0 at length $T$ minutes, similar to previous experiments~\cite{seeley91,granovskiy12}. Before analyzing the temporal dynamics $u(t)$ of the swarm in response to food quality switches, we study equilibria and their stability to determine different swarm interactions impact foraging consensus and the rate at which it is approached.

\subsection{Equilibrium and linear stability analysis} \label{equilanaz}
At any given time $t$, the dynamics of Eq.~(\ref{singledyn}) are determined by the food quality function $\alpha(t)$ values at $t$ and $t - \tau$. In the time interval $ t \in [0,\tau] $, $\alpha(t) = \balpha$ and $\alpha(t - \tau) = 0$ equilibria of Eq.~(\ref{singledyn}) are solutions to
\begin{align*}
0 &= (1-u)(\balpha + \beta u) - \gamma u - \rho \balpha u^2, 
\end{align*}
which can be solved using the quadratic formula
\begin{align}
\bu_{\pm}^1 := \frac{1}{2} \left[ {\mc B} \pm \sqrt{{\mc D}} \right], \hspace{10mm} {\mc B} = \frac{\beta - \gamma - \balpha}{\beta + \rho \balpha}, \hspace{8mm} {\mc D} = {\mc B}^2 + \frac{4 \balpha}{\beta + \rho \balpha},  \label{u1eq}
\end{align}
with linear stability given by the eigenvalues
\begin{align*}
\lambda_{\pm}^1 = \mp \sqrt{(\beta - \bar{\alpha} - \gamma)^2 + 4 (\beta + \rho\balpha ) \bar{\alpha}},
\end{align*}
so the positive equilibrium $\bar{u}_+^1$ is stable and the negative (extraneous) equilibrium $\bar{u}_-^1$ is unstable.
On $ t \in [\tau,T] $, $\alpha(t) = \alpha(t- \tau) = \balpha$ the equilibrium equation
\begin{align*}
0 &= (1-u)(\balpha + \beta u) - \gamma u .
\end{align*}
has solutions and eigenvalues 
\begin{align*}
\bar{u}_{\pm}^2 = \frac{\beta - \bar{\alpha} - \gamma \pm \sqrt{(\beta - \bar{\alpha} - \gamma)^2 + 4 \beta \bar{\alpha}}}{2 \beta}, \ \ \ \ \
\lambda_{\pm}^2 = \mp \sqrt{(\beta - \bar{\alpha} - \gamma)^2 + 4 \beta \bar{\alpha}}.
\end{align*}
Again, the positive equilibrium $\bar{u}_+^2$ is stable and the negative equilibrium $\bar{u}_-^2$ is unstable.
On $t \in [T, T+ \tau)$, $\alpha(t) = 0$ and $\alpha(t- \tau) = \bar{\alpha}$, equilibria satisfy $0 = (1-u) \beta u - \gamma u$, so
\begin{align*}
&\bar{u}_0^3 = 0, \quad \bar{u}_1^3 = \frac{\beta - \gamma}{\beta},
\end{align*}
and on $t \in [T+ \tau, 2T)$, $\alpha(t)= \alpha(t - \tau)= 0$, so $0 = (1-u) \beta u - \gamma u - \rho \bar{\alpha} u^2$ and
\begin{align*}
&\bar{u}_0^4 = 0, \quad \bar{u}_1^4 = \frac{\beta - \gamma}{\beta + \rho \bar{\alpha}}.
\end{align*}
Both pairs of equilibria have associated eigenvalues
\begin{align*}
\lambda_0 = \beta - \gamma, \ \ \ \ \ \lambda_1 = \gamma - \beta,
\end{align*}
so the zero equilibria $\bar{u}_0^3 = \bar{u}_0^4 = 0$ are stable when $\gamma > \beta$ and the nonzero equilibria $\bar{u}_1^3$ and $\bar{u}_1^4$ are positive and stable when $\beta > \gamma$. Thus, to ensure no bees continue foraging when there is no food, abandonment $\gamma$ should be stronger than recruitment $\beta$.

We deem $\bar{u} : = \bar{u}_+^2$ the {\em consensus} level, as it is the upper limit on the fraction of the swarm foraging the feeder, when it supplies food. The eigenvalues $\lambda_{\rm on} : = \lambda_+^2$ and $\lambda_{\rm off} : = \lambda_0^4$ define the {\em adaptivity} of the swarm, or the rates of arrival to/departure from the feeder when it does/does not supply food.


\subsection{Periodically forced swarm foraging}
\label{evolnmean}

Long term periodic solutions to Eq.~(\ref{singledyn}) result from switching the food quality $\alpha(t)$ between $\bar{\alpha}$ and $0$ every $T$ minutes. These are obtained by solving Eq.~(\ref{singledyn}) iteratively using separation of variables. For example, when $\alpha (t) \equiv \bar{\alpha}$ and $\alpha(t-\tau) \equiv 0$ we can separate variables and factor the resulting fraction:
\begin{align*}
\frac{ \d u}{u - \bar{u}_+} - \frac{ \d u}{u - \bu_-}  = - (\beta + \rho \balpha) \sqrt{{\mc D}} \d t,  
\end{align*}
where ${\mc D}$ is defined in Eq.~(\ref{u1eq}). Integrating,
 isolating $u$, and applying $u(0) = u_0$, we find
\begin{align} \label{u1soln}
u(t) = \frac{\bu_+ ( u_0 - \bu_-) - \bu_-(u_0 - \bu_+) \e^{- (\beta + \rho \balpha) \sqrt{\mc D} t}}{u_0 - \bu_- - (u_0 - \bu_+) \e^{- (\beta + \rho \balpha)\sqrt{\mc D}t}}, 
\end{align}
consistent with our equilibrium analysis showing $\lim_{t \to \infty} u(t) = \bu = \bu_+^2$. Now, taking $\alpha (t) \equiv \bar{\alpha}$ on $t \in [2nT, (2n+1)T)$ for $n=0,1,2,3,...$ and $\bar{\alpha} \equiv 0$ otherwise, we will have
\begin{align}
\dot{u} = (1-u)(\alpha(t) + \beta u) - \gamma u - R(t) u^2, \label{simpsing}
\end{align}
where $R(t) \equiv \rho \bar{\alpha}$ for $t \in [2nT + \tau, (2n+1)T + \tau)$ and $R(t) \equiv 0$ otherwise.
The periodic solution to Eq.~(\ref{simpsing}) can be derived self-consistently by starting with an unknown initial condition $u(0) = u_0$, and then requiring $u(2T) = u_0$. Thus, within $t \in [0,\tau)$, we have the solution given by Eq.~(\ref{u1soln}), and
\begin{align}  \label{u1eqn}
u_1 : = u(\tau) = \frac{\bu_+ ( u_0 - \bu_-) - \bu_-(u_0 - \bu_+) \e^{- (\beta + \rho \balpha) \sqrt{\mc D} \tau}}{u_0 - \bu_- - (u_0 - \bu_+) \e^{- (\beta + \rho \balpha)\sqrt{\mc D} \tau }}. 
\end{align}
At $t = \tau$, self-inhibition vanishes and the solution is a special case of Eq.~(\ref{u1soln}) for which $\rho = 0$. Thus, we can solve Eq.~(\ref{singledyn}) with $u(\tau) = u_1$ as an initial condition and write for $t \in [\tau, T)$:
\begin{align} \label{u2soln}
u(t) = \frac{\bu_+ ( u_1 - \bu_-) - \bu_-(u_1 - \bu_+) \e^{- \beta \sqrt{\mc D} t}}{u_1 - \bu_- - (u_1 - \bu_+) \e^{- \beta \sqrt{\mc D}t}},
\end{align}
so that at $t = T$, we have
\begin{align} \label{u2eqn}
u_2 : = u(T) = \frac{\bu_+ ( u_1 - \bu_-) - \bu_-(u_1 - \bu_+) \e^{- \beta \sqrt{\mc D} (T-\tau)}}{u_1 - \bu_- - (u_1 - \bu_+) \e^{- \beta \sqrt{\mc D}(T -\tau)}}.
\end{align} 
Beyond $t=T$, the dynamics is governed by a special case of Eq.~(\ref{u2soln}) for which $(\bu_+, \bu_-) = (1-\frac{\gamma}{\beta} , 0)$ if $\beta > \gamma$ and  $(\bu_+, \bu_-) = (0,1-\frac{\gamma}{\beta} )$ if $\beta < \gamma$, so on $t \in [T, T + \tau)$:
\begin{align*}
u(t) = \frac{u_2( \beta - \gamma )}{\beta u_2 - (\beta u_2 + \gamma - \beta) \e^{(\gamma - \beta ) t}},
\end{align*}
for $\beta \neq \gamma$, and the limit as $\gamma \to \beta$ is $u(t) = \frac{u_2}{1 + u_2 \beta t}$, which can both be evaluated at $t = T+ \tau$ to yield,
\begin{align} \label{u3eqn}
u_3 : = u(T+\tau) = \left\{ \begin{array}{cc} 
\frac{u_2( \beta - \gamma )}{\beta u_2 - (\beta u_2 + \gamma - \beta) \e^{(\gamma - \beta )\tau}}  & : \  \beta \neq \gamma \\
\frac{u_2}{1 + u_2 \beta \tau } & : \  \beta = \gamma 
 \end{array} \right.
\end{align}

At $t = T+\tau$, self inhibition returns since $\alpha (t- \tau) \equiv 0$, increases the negative feedback acting on foragers.
The long term steady state is determined by the balance of abandonment and recruitment: $(\bu_+, \bu_-) = (\frac{\beta - \gamma}{\beta + \rho \balpha} , 0)$ if $\beta > \gamma$ and  $(\bu_+, \bu_-) = (0,\frac{\beta - \gamma}{\beta + \rho \balpha} )$ if $\beta < \gamma$. Thus,
\begin{align*} 
u(t) = \frac{u_3( \beta - \gamma )}{ (\beta + \rho \balpha) u_3 - ((\beta + \rho \balpha) u_3 + \gamma - \beta) \e^{(\gamma - \beta ) t}}, 
\end{align*}
for $\beta \neq \gamma$, and in the limit $\beta \to \gamma$, $u(t) = \frac{u_3}{1 + u_3 (\beta + \rho \balpha) t}$. Both expressions can be evaluated at $t = 2T$, and self-consistency of the periodic solution requires $u_4 \equiv u_0$,
\begin{align} \label{u4eqn}
u_0 = u_4 : = u(2T) = \left\{ \begin{array}{cc} 
\frac{u_3( \beta - \gamma )}{ (\beta + \rho \balpha) u_3 - ((\beta + \rho \balpha) u_3 + \gamma - \beta) \e^{(\gamma - \beta ) (T-\tau) }} & : \  \beta \neq \gamma \\
\frac{u_3}{1 + u_3 (\beta + \rho \balpha) (T - \tau)} & : \  \beta = \gamma 
\end{array} \right.
\end{align}
Eqs.~(\ref{u1eqn}), (\ref{u2eqn}), (\ref{u3eqn}), and (\ref{u4eqn}) can be solved explicitly for $(u_0, u_1, u_2, u_3)$, although the expressions are quite cumbersome, so we omit them here. These analytic solution techniques were used to generate the foraging fraction trajectories plotted in Fig.~\ref{fig2}d and to identify model parameter that optimize the RR $J$ in different environments (plotted in Fig.~\ref{fig2}e) as we now describe.

\subsection{Optimizing reward rate over strategy sets}
\label{optimone}
We optimized the reward rate (RR) of the swarm foraging a single switching feeder by restricting the strategies to a discrete set of interaction parameter values. The RR in large regions of parameter space was relatively flat since it involves the sum of several exponentially small terms. To avoid spurious convergence, we focused on each parameter's relevant order of magnitude which led to the highest long term RR. For a given environment $(\alpha, T)$, we identified the combination of interaction parameter values from the set $(\beta,\gamma,\rho) \in \{ 0.01,0.1,1,10\}^3$ (in 1/min) yielding the highest RR $J$ computed from Eq.~(\ref{rr1}). Bounds on interaction parameters were imposed so that a swarm could not completely dispense with any interaction or feedback mechanism or strengthen any to be arbitrarily rapid. This was performed over a mesh of environmental parameters $\bar{\alpha} \in [0.5,20]$ (at $\Delta \bar{\alpha} = 0.1$ steps) and $T \in [1,200]$ (at $\Delta T = 1$ minute). We found that $\beta = 0.01 \text{ min}^{-1}$ was optimal across all environment types, but that $\gamma$ and $\rho$ varied in strength dependent on the environmental conditions (See Fig.~\ref{fig2}e).

\subsection{Linear approximation of the periodic solution and reward rate}		\label{linearappone}

The RR Eq.~(\ref{rr1}) for the single feeder can be estimated by linearly approximating the swarm dynamics using results from our equilibrium analysis. Assuming the interval $T$ and between feeder quality switches ($\alpha: \bar{\alpha} \mapsto 0; \alpha: 0 \mapsto \bar{\alpha}$) and the delay $\tau$ are large, the swarm will nearly equilibrate before each switch, suggesting the following linear approximation of the foraging fraction:
\begin{align*}
u(t) = \left\{ \begin{array}{cc} 
\bar{u}^{1}  + \e^{- \lambda^{1} t} (\bar{u}^{4}  -\bar{u}^{1} ), & t \in [0,\tau]   \\
\bar{u}^{2}  + \e^{- \lambda^{2} (t-\tau)} (\bar{u}^{1}  -\bar{u}^{2} ), &  t \in [\tau,T]  \\
\bar{u}^{3}  + \e^{- \lambda^{3} (t-T)} (\bar{u}^{2}  - \bar{u}^{3} ), &  t \in [T,T+\tau]  \\
\bar{u}^{4}  + \e^{- \lambda^{4} (t-T-\tau)} (\bar{u}^{3} - \bar{u}^{4} ), &  t \in [T+\tau,2T] .
\end{array} \right.
\end{align*}
where $\bar{u}^{i}$ are the stable equilibria and $\bar{u}^3 = \bar{u}^{4} = 0$ when $\beta < \gamma$. Considering this case, we can compute the RR using the single feeder version of Eq.~(\ref{rr2}) in the long time limit
\begin{align*}
J =& \frac{1}{2T} \int_0^{2T} u(t) (\alpha (t) - c) \d t \\
=& \frac{\bar{\alpha} - c}{2T} \int_0^\tau \bar{u}^1(1 -  \e^{- \lambda^1 t} ) \d t + \frac{\bar{\alpha} - c}{2T} \int_0^{T-\tau} (\bar{u}^2 +  \e^{- \lambda^2 t} (\bar{u}^{1}  -\bar{u}^{2} )) \d t - \frac{c}{2T} \int_0^\tau \bar{u}^{2} \e^{- \lambda^3 t} \d t \\
=& \frac{\bar{\alpha} - c}{2T} \left( \bar{u}^1 \tau - \bar{u}^1 \frac{1- \e^{- \lambda^1 \tau} }{\lambda^1} \right)+ \frac{\bar{\alpha} - c}{2T} \left( \bar{u}^2 (T-\tau) + \frac{\bar{u}^{1}  -\bar{u}^{2}}{\lambda^2} (1- \e^{- \lambda^2 (T-\tau)} ) \right) - \frac{c}{2T} \left( \frac{\bar{u}^{2}}{\lambda^3} (1- \e^{- \lambda^3 \tau} ) \right).
\end{align*}
In the the long interval $\lim_{T \to \infty}$ and short delay $\lim_{\tau \to 0}$ (omitting the intermediate delay equilibria) limits, we can simplify the expression as
\begin{align*}
J &=  \frac{\bar{u}}{2} \left[ (\bar{\alpha} - c) \left( 1 - \frac{1 - \e^{- \lambda_{\rm on} T}}{\lambda_{\rm on} T} \right)- c \frac{1 - \e^{- \lambda_{\rm off} T}}{\lambda_{\rm off} T}  \right],
\end{align*}
where $\bar{u} = \bar{u}^2$, $\lambda_{\rm on} = \lambda^2 = \lambda_+^2$ and $\lambda_{\rm off} = \lambda^4 = \lambda_0$, as written in Eq.~(\ref{linJone}). For the specific case in which $\bar{\alpha} = 2$ and $c = 1$, we can write this more cleanly as
\begin{align*}
J(\alpha(t), \beta, \gamma, \rho) =  \frac{\bar{u}}{2} \left[ \left( 1 - \frac{1 - \e^{- \lambda_{\rm on} T}}{\lambda_{\rm on} T} \right)-  \frac{1 - \e^{- \lambda_{\rm off} T}}{\lambda_{\rm off} T}  \right].
\end{align*}
Clearly, increasing consensus ($\bar{u}$) and adaptivity ($\lambda_{\rm on/off}$) increases the swarm RR.

As $\beta \to 0$, $\lambda_{\rm off} = - \gamma$, $\bar{u} = 2/[ 2 + \gamma ]$, with $\lambda_{\rm on} = - (2 + \gamma)$. Increasing the rate of abandonment $\gamma$ decreases consensus $\bar{u}$ but will increase the adaptivity of the swarm as both $\lambda_{\rm off} = - \gamma$ and $\lambda_{\rm off} = - (\gamma +2)$ increase in amplitude. Optimizing the RR then requires balancing these two effects. Identifying the $\gamma$ value that maximizes the RR can then done by finding the maximum of
\begin{align*}
J(\alpha(t), 0, \gamma, \rho) = \frac{1}{2 + \gamma} \left[ \left( 1 - \frac{1 - \e^{- (2 + \gamma) T}}{(2 + \gamma)T} \right)-  \frac{1 - \e^{- \gamma T}}{\gamma T}  \right],
\end{align*}
given by the $\gamma$ solving $\pd_{\gamma} J(\alpha(t), 0, \gamma, \rho)  = 0$. This analysis can be extended to consider the solutions to the full nonlinear equations, but the general trends are the same. Increasing negative feedback will tend to limit consensus while making the swarm more adaptive to change.

\section{Dynamics of swarms foraging at two switching feeders}

Here we provide more details and analysis on our swarm model Eq.~(\ref{2siteswarm}) foraging between two feeders. As in the single feeder model, we can leverage equilibrium, stability, and linearization to better understand the impact of model tuning on the RRs of the foraging swarm.

\subsection{Forms of social inhibitions} \label{socialin}

Here, we provide detailed descriptions of the dynamical models associated with each type of social inhibition used in the model of a swarm foraging two feeders, as generalized in Eq.~(\ref{2siteswarm}). In the main text, we simply indicate the general form of social inhibition with the function ${\mc S}(u_A, u_B)$, but we provide the functional form for these interactions in the descriptions below. \\
\vspace{-3mm}

\noindent
{\em Direct Switching Model:} A bee committed to a feeder inhibits bees with opposing opinions by causing them to switch the feeder to which they are committed:
\begin{align*} 
\dot{u_A} = & (1-u_A-u_B)(\alpha_A(t) + \beta u_A) - \gamma u_A - \rho(\alpha_B(t-\tau) - \alpha_A(t-\tau)) u_Au_B,  \\
\dot{u_B} = & (1-u_A-u_B)(\alpha_B(t) + \beta u_B) - \gamma u_B - \rho(\alpha_A(t-\tau) - \alpha_B(t-\tau))u_Au_B, 
\end{align*}
where $\tau$ (in minutes) indicates the time delay required for the strength of the direct switching signal (based on detected food quality) to update following a switch in the food quality. Notice that the social inhibition terms in either evolution equation ($u_A$ or $u_B$) will necessarily be of opposite sign since each is the negative of the other and $\alpha_A(t) \neq \alpha_B(t)$ for all $t>0$. \\
\vspace{-3mm}

\noindent
{\em Indiscriminate Stop Signal Model:} A bee committed to a feeder indiscriminately inhibits bees committed to either feeder, affecting both bees committed to the same feeder and those committed to a different feeder, and causes them become uncommitted:
\begin{align*}
\dot{u_A} = & (1-u_A-u_B)(\alpha_A(t) + \beta u_A) - \gamma u_A - \frac{1}{2}\rho(\alpha_A(t-\tau)u_A^2 + \alpha_B(t-\tau) u_Au_B),  \\
\dot{u_B} = & (1-u_A-u_B)(\alpha_B(t) + \beta u_B) - \gamma u_B - \frac{1}{2}\rho(\alpha_B(t-\tau)u_B^2 + \alpha_A(t-\tau) u_Au_B),
\end{align*}
where $\tau$ (in minutes) is the time delay required for food quality switch detection as in the direct switching model. This form of social inhibition will always lead to negative feedback to both populations as long as $\rho >0$. \\
\vspace{-3mm}

\noindent
{\em Self-Inhibition Model:} A bee committed to one feeder inhibits bees committed to the same feeder, causing them to become uncommitted:
\begin{align*}
\dot{u_A} = & (1-u_A-u_B)(\alpha_A(t) + \beta u_A) - \gamma u_A - \rho(\bar{\alpha} - \alpha_A(t-\tau)) u_A^2,  \\
\dot{u_B} = & (1-u_A-u_B)(\alpha_B(t) + \beta u_B) - \gamma u_B - \rho(\bar{\alpha} - \alpha_B(t-\tau))u_B^2,
\end{align*}
where $\tau$ is the time delay. Self-inhibition is only active in the population of foragers  for which the swarm detects there is less than the maximum supply of food available ($\alpha_{A,B}(t-\tau) < \bar{\alpha}$). \\
\vspace{-3mm}

\noindent
{\em Discriminate Stop Signal:} A bee committed to a feeder inhibits bees committed to different feeders, causing them to become uncommitted:
\begin{align*} 
\dot{u_A} = & (1-u_A-u_B)(\alpha_A(t) + \beta u_A) - \gamma u_A - \rho \alpha_B(t-\tau) u_Au_B,  \\
\dot{u_B} = & (1-u_A-u_B)(\alpha_B(t) + \beta u_B) - \gamma u_B - \rho \alpha_A(t-\tau) u_Au_B, 
\end{align*}
where $\tau$ is the time delay. The strength of the stop signal varies with the detected quality of each feeder.

\subsection{Equilibria and linear stability} \label{twolinstab}

We determined linear approximations of periodic solutions to the full nonlinear model Eq.~(\ref{2siteswarm}) by studying the equilibria and linear stability properties of the full system. For any $t$, the dynamics of Eq.~(\ref{2siteswarm}) is governed by the piecewise constant values of the food quality functions denoted $\bar{\alpha}_{A,B}^t = \alpha_{A,B}(t)$ and $\bar{\alpha}_{A,B}^{\tau} = \alpha_{A,B}(t - \tau)$ as the actual current and delay-observed values so that:
\begin{subequations} \label{2dfp}
\begin{align}
0 &= (1-\bu_A - \bu_B)( \balpha_A^t + \beta \bu_A) - \gamma \bu_A - {\mc S}(\bu_A, \bu_B; \rho, \balpha_A^{\tau}, \balpha_B^{\tau}), \\
0 &= (1-\bu_A - \bu_B)( \balpha_B^t + \beta \bu_B) - \gamma \bu_B - {\mc S}(\bu_B, \bu_A;  \rho, \balpha_B^{\tau}, \balpha_A^{\tau}). 
\end{align}
\end{subequations}
where ${\mc S}(x,y; \alpha_x, \alpha_y)$ is the nonlinear function describing inhibitory social interactions, parameterized by the strength $\rho$ and delayed quality observations $\balpha_{A,B}^{\tau}$ (See Section~\ref{socialin} for exact forms).  Since Eq.~(\ref{2dfp}) is autonomous for piecewise time intervals, equilibria can be defined on each interval~\cite{bernardo08}. Eq.~(\ref{2dfp}) can be explicitly solved using the quartic equation for all models and time intervals, but the expressions for $\bu_{A,B}$ are unwieldy, so we do not write them here.

Linear stability was classified using the eigenvalues $\lambda_{\pm} = \frac{1}{2} \left[ {\rm Tr} ({\mathcal D}) \pm \sqrt{{\rm Tr}({\mathcal D})^2 - 4 {\rm det}({\mathcal D})} \right]$ of the Jacobian about fixed points $(u_A, u_B) = (\bu_A, \bu_B)$:
\begin{eqnarray*}  \begin{aligned}
\mathcal{D} = \begin{bmatrix} 		
     {- \balpha + \beta(1- \bar{u}_B - 2 \bar{u}_A) - \gamma - \partial_{u_A} {\mc S}(\bar{u}_A, \bar{u}_B)} & { - \balpha - \beta \bar{u}_A - \partial_{u_B} {\mc S}(\bar{u}_A, \bar{u}_B)} \\
      {- \frac{\balpha}{2} - \beta \bar{u}_B - \partial_{u_A} {\mc S}(\bar{u}_B, \bar{u}_A)} &  { - \frac{\balpha}{2} + \beta(1 - \bar{u}_A - 2 \bar{u}_B) - \gamma - \partial_{u_B} {\mc S}(\bar{u}_B, \bar{u}_A)} 
\end{bmatrix}.
\end{aligned} \end{eqnarray*}
Specific cases are stable nodes (with two negative real eigenvalues, $\lambda_{\pm} <0$) and saddles (with one negative/one positive real eigenvalue, $\lambda_{\pm} \gtrless 0$) as illustrated Fig.~\ref{fig3}. Direct switching, self inhibition and indiscriminate stop signaling yield monostable behavior -- a phase space only containing a single stable node (Fig.~\ref{fig3}e) -- and the majority of bees forage at the high yielding feeder ($\bu_A > \bu_B$).

Discriminate stop signaling model can generate bistability for weak abandonment $\gamma$ and strong recruitment $\beta$ and stop-signaling $\rho$. In this case, the phase space is occupied by two stable nodes separated by a saddle point (Fig.~\ref{fig3}f). In this case, consensus is lower in some phases of the foraging cycle, since discriminate stop-signaling prevents switching between feeders.

\subsection{Optimizing reward rate over strategy sets}	\label{optimtwo}

As in the one feeder case, identified the set $(\beta,\gamma,\rho) \in \{ 0.001,0.1,1,10\}^3$  (min$^{-1}$) yielding the highest RR from Eq.~(\ref{rr2}) in each environment $(\bar{\alpha},T)$. For each parameterized form of social inhibition, we numerically found periodic solutions to Eq.~(\ref{2siteswarm}) taking $\alpha_A = \bar{\alpha}$ and $\alpha_B = \frac{\balpha}{2}$ initially and flipping the feeder qualities every $T$ minutes. This was performed over a mesh of environmental parameters $\bar{\alpha} \in [0.5,20]$ (at $\Delta \bar{\alpha} = 0.1$ steps) and $T \in [1,200]$ (at $\Delta T = 1$ minute). See Fig~\ref{fig5} for direct switching and Figs. \ref{fig8} and \ref{fig9} for discriminate and indiscriminate stop signaling model respectively. For the self inhibition model, the optimal strategy is low abandonment ($\gamma = 0.01 \text{ min}^{-1}$) with high recruitment and social inhibition ($\beta = \rho = 10 \text{ min}^{-1}$). The maximum RR is plotted for each of the four models in a given environmental condition (food quality $\balpha$ and switching period $T$) in Fig~\ref{fig4}. 

\subsection{Linear approximation of the periodic solution} \label{linearapptwo}

To compute consensus and adaptivity, we derived a linear approximation to the period solution in the two feeder case. Feeder qualities started with $\alpha_A(t) = \bar{\alpha}$ and $\alpha_B(t) = \frac{\balpha}{2}$ and switched every $T$ minutes. Assuming $T$ large, the swarm will equilibrate between condition switching, yielding the following estimate of the $u_A(t)$ part of the periodic solution:
\begin{align*}
u_A(t) = \left\{ \begin{array}{cc} 
\bar{u}^{1}  + \e^{- \lambda^{1} t} (\bar{u}^{4}  -\bar{u}^{1} ), & t \in [0,\tau],   \\
\bar{u}^{2}  + \e^{- \lambda^{2} (t-\tau)} (\bar{u}^{1} -\bar{u}^{2} ), &  t \in [\tau,T],  \\
\bar{u}^{3}  + \e^{- \lambda^{3} (t-T)} (\bar{u}^{2}  - \bar{u}^{3} ), &  t \in [T,T+\tau],  \\
\bar{u}^{4}  + \e^{- \lambda^{4} (t-T-\tau)} (\bar{u}^{3}- \bar{u}^{4} ), &  t \in [T+\tau,2T],
\end{array} \right.
\end{align*}
where $\bar{u}^{i}$ are stable equilibria in each time interval and $\lambda^j$ are the least negative associated eigenvalues determining the decay rate to the fixed point. There is a similar expression for the opposing feeder population, $u_B(t) = u_A(t+T)$. This implies $\lambda^1 = \lambda^3$ and $\lambda^2 = \lambda^4$. In the long time limit, the RR Eq.~(\ref{rr2}) can be computed:
\begin{align*}
J =& \frac{1}{T_f} \int_0^{T_f} \left[ u_A(t) \cdot (\alpha_A (t) - c) + u_B(t) \cdot (\alpha_B (t) - c) \right] \d t \\
=& \frac{\bar{\alpha} - c}{2T} \int_0^\tau (\bar{u}^1 +  \e^{- \lambda^1 t} (\bar{u}^{4}  -\bar{u}^{1} )) \d t + \frac{\bar{\alpha}/2-c}{2T}  \int_0^\tau (\bar{u}^3 +  \e^{- \lambda^1 t} (\bar{u}^{2}  -\bar{u}^{3} )) \d t \\
& + \frac{\balpha - c}{2T} \int_0^{T-\tau} (\bar{u}^2 +  \e^{- \lambda^2 t} (\bar{u}^{1}  -\bar{u}^{2} )) \d t +  \frac{\bar{\alpha}/2-c}{2T}  \int_0^{T - \tau} (\bar{u}^4 +  \e^{- \lambda^2 t} (\bar{u}^{3}  -\bar{u}^{4} )) \d t \\
& + \frac{\bar{\alpha}/2-c}{2T} \int_0^\tau ( \bar{u}^3  + \e^{- \lambda^1 t} (\bar{u}^{2}  - \bar{u}^{3} ))\d t  + \frac{\balpha - c}{2T} \int_0^{\tau}( \bar{u}^1  + \e^{- \lambda^1 t} (\bar{u}^{4}  - \bar{u}^{1} ))\d t \\
& +  \frac{\bar{\alpha}/2-c}{2T}  \int_0^{T-\tau} ( \bar{u}^4  + \e^{- \lambda^2 t} (\bar{u}^{3}  - \bar{u}^{4} ))\d t + \frac{\balpha - c}{2T} \int_0^{T-\tau} ( \bar{u}^2  + \e^{- \lambda^2 t} (\bar{u}^{1}  - \bar{u}^{2} ))\d t \\
=& \frac{\bar{\alpha} - c}{T} \left[ \left( \bar{u}^1 \tau + \frac{\bar{u}^{4}  -\bar{u}^{1}}{\lambda^1} (1- \e^{- \lambda^1 \tau} ) \right)+ \left( \bar{u}^2 (T-\tau) + \frac{\bar{u}^{1} -\bar{u}^{2}}{\lambda^2} (1- \e^{- \lambda^2 (T-\tau)} ) \right) \right] \\
& + \frac{\bar{\alpha}/2 - c}{T} \left[ \left( \bar{u}^3 \tau + \frac{\bar{u}^{2}  -\bar{u}^{3}}{\lambda^3} (1- \e^{- \lambda^3 \tau} ) \right)  + \left( \bar{u}^4 (T-\tau) + \frac{\bar{u}^{3}  -\bar{u}^{4}}{\lambda^4} (1- \e^{- \lambda^4(T-\tau)} ) \right)\right] .
\end{align*}
Considering the limit of long time intervals $\lim_{T \to \infty}$ and short delays $\lim_{\tau \to 0}$ and the case in which $\bar{u}^2 + \bar{u}^4 \approx 1$ (no uncommitted bees in the long time limit) we further simplify the expression:
\begin{align*}
J &=  (\balpha - c) \left[ \bar{u}^2 + (1 - 2 \bar{u}^2) \frac{1 - \e^{- \lambda^2 T}}{\lambda^2 T} \right] + (\balpha/2 - c) \left[ 1-\bar{u}^2 + (2 \bar{u}^2-1) \frac{1 - \e^{- \lambda^4 T}}{\lambda^4 T}  \right].
\end{align*}
For the specific case in which $c = \frac{\balpha}{2}$, we remove the superscripts so $\bar{u} = \bar{u}^2$ and $\lambda = \lambda^2$:
\begin{align*}
J  = \frac{\bar{\alpha}}{2} \left( \bar{u} + (1- 2\bar{u})\frac{1 - \e^{- \lambda T}}{\lambda T} \right).
\end{align*}
The gradient of the RR along $\bar{u}$ and $\lambda$ can then be computed as:
\begin{align*}
\partial_{\bar{u}}J &= \frac{\balpha}{2} \left( 1 - 2 \frac{1 - \e^{- \lambda T}}{\lambda T} \right), \ \ \ \ \ \ \ \ \partial_{\lambda}J = (2 \bar{u} - 1)\frac{\balpha \e^{-\lambda T}}{2 \lambda^2 T} (\e^{\lambda T} - \lambda T - 1),
\end{align*}
showing $J$ is increasing in $\bar{u}$ as long as $\lambda T > 1.594$ and increasing in $\lambda$ as long as $\bar{u}>0.5$.

\section{Supplemental figures and table}

\begin{table}[t!]
\begin{center}
\caption{Model parameters for single feeder Eq.~(\ref{singledyn}) and two feeder Eq.~(\ref{2siteswarm}) swarm foraging models.}
\label{table1}
	\begin{tabular}{|c c c c|} \hline 
   {parameter} & {description} & {numerical range} & {citation} \\[1ex] \hline
   {$\bar{\alpha}$} & {quality of food source}	&	{$[0.5,20]$ M(mol/l)} & {\cite{seeley00,granovskiy12}} \\[1ex] 
   {$\beta$} & {recruitment rate}	& {$\mathcal{O}(10^{-1} - 10^{1})$ min$^{-1}$} & {\cite{sumpter03}, \cite{seeley12}~supplement} \\[1ex] 
   {$\gamma$} & {abandonment rate}	&{$\mathcal{O}(10^{-2} - 10^{1})$ min$^{-1}$} & {\cite{seeley12}~supplement} \\[1ex] 
   {$\rho$} & {rate of social inhibition}	& {$\mathcal{O}(10^{-1} - 10^{1})$ min$^{-1}$} & {\cite{seeley12}~supplement} \\[1ex] 
   {$T$} & {period of environment switch}	& {$1-200$ min} & {\cite{granovskiy12}} \\[1ex]
   {$\tau$} & {time delay for switch to be sensed}	&{$ 0.1 \cdot T$ min} & {--} \\[1ex]
   {$c$} &{cost of foraging}	& { $ \frac{\bar{\alpha}}{2} $ } & {--} \\[1ex] \hline

\end{tabular}
\end{center}

\end{table}

\begin{figure}[b!]
\begin{center} \includegraphics[width=.4\linewidth]{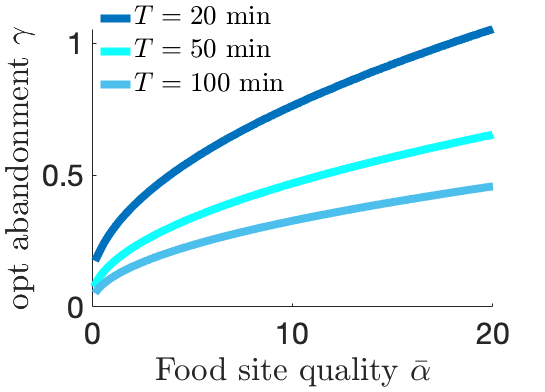} \end{center}
\caption{Reward rate, Eq.~(\ref{rr1}), maximizing values of abandonment ($\gamma$) parameter for a given food quality ($\alpha$) and switching period ($T$) in the single feeder model. Swarm can maximize the reward $J$ by calibrating the level of abandonment with the switching rate and feeder quality, discounting faster as the environment changes more quickly.} 	\label{fig7}
\end{figure}

\subsection{Matching abandonment rate to switching rate in a single dynamic feeder} \label{abandon}
Considering the single feeder foraging swarm model without nonlinear negative feedback ($\rho = 0$ so delays $\tau$ are irrelevant), we can explicitly compute the RR $J$ as a function of other parameters. We found that the best strategies do not utilize recruitment ($\beta = 0$), so the abandonment rate $\gamma$ is the only parameter that needs to be tuned with the environmental switching time $T$ and food quality $\bar{\alpha}$.

Thus, Eq.~(\ref{singledyn}) is linear and so the linear approximation of the periodic solution is exact, described by
\begin{align*}
u(t) = \left\{ \begin{array}{cc} A + (u_0 - A) \e^{-(\balpha + \gamma) t}, & t \in [0,T), \\
u_1 \e^{-\gamma (t-T)}, & t \in [T,2 T) \end{array} \right.
\end{align*}
where $u_0 =A (1 - \e^{- (\balpha + \gamma) T})/(\e^{\gamma T} - \e^{- (\balpha + \gamma) T}) $ and $u_1 =A (\e^{\gamma T} - \e^{- \balpha T})/(\e^{\gamma T} - \e^{- (\balpha + \gamma) T})  $. As such, we can explicitly compute the RR Eq.~(\ref{rr1}),
\begin{align*}
J = \frac{\balpha - c}{2T} \left( AT + \frac{u_0 - A}{\balpha + \gamma} ( 1- \e^{- (\balpha + \gamma)T}) \right) - \frac{c}{2 T} \left( \frac{A}{\gamma } ( 1 - \e^{- \gamma T}) \right),
\end{align*}
determining the maximum with respect to the abandonment rate $\gamma$ by solving $\partial_{\gamma} J = 0$ (Fig.~\ref{fig7}). The optimal abandonment rate $\gamma$ increases with the food site quality and decreases with the switching time $T$ of the environment. Thus, the negative feedback process should adapt to the dynamics of the environment, and discounting can be more rapid when the evidence for feeder quality is stronger.

\begin{figure}[t]
\begin{center}
{\includegraphics[width = \linewidth]{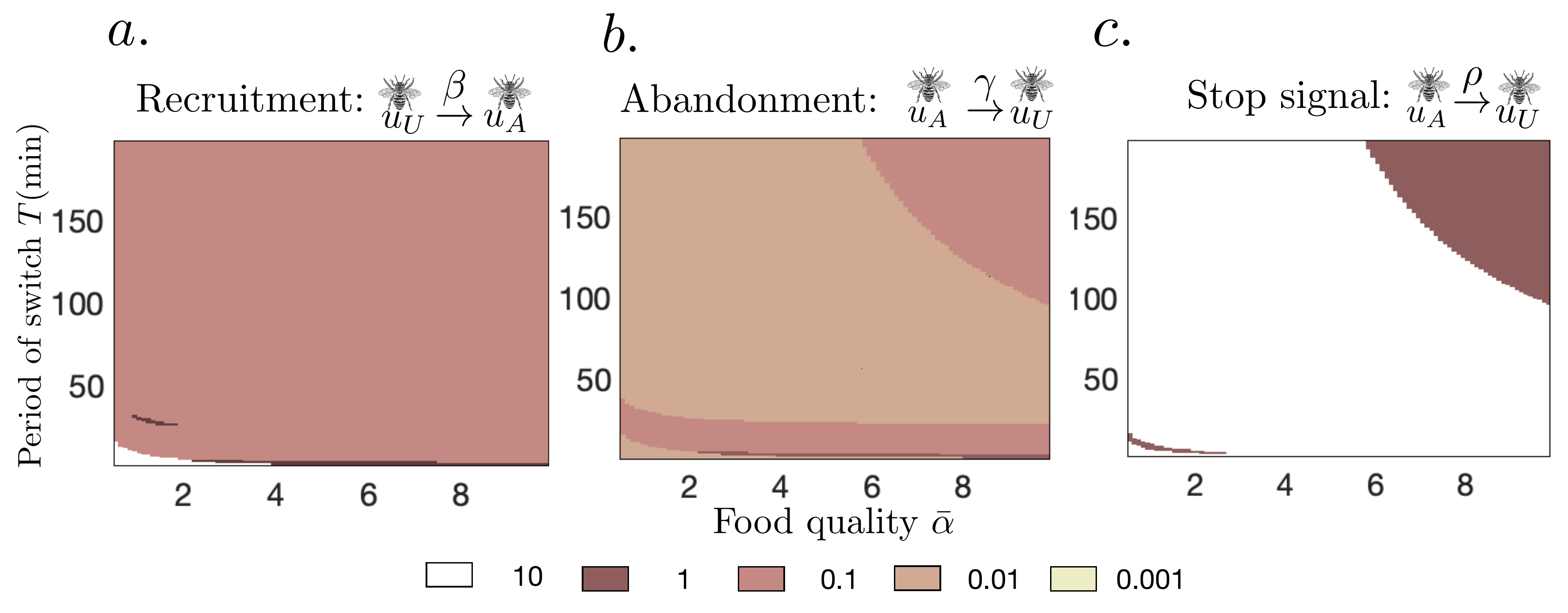}}
\caption{Tuning ({\bf a}) recruitment $\beta$;  ({\bf b}) abandonment $\gamma$; and  ({\bf c}) social inhibition $\rho$ to maximize the reward rate (RR), Eq.~(\ref{rr2}) in the discriminate stop signaling model. ({\bf a}) Recruitment and ({\bf b}) abandonment should be made weak whereas ({\bf c}) social inhibition should be made strong except for in slow (high $T$) and high quality $\bar{\alpha}$ environments.}	\label{fig8}
\end{center}
\end{figure}

\begin{figure}[t!]
\begin{center}
{\includegraphics[width = 1\linewidth]{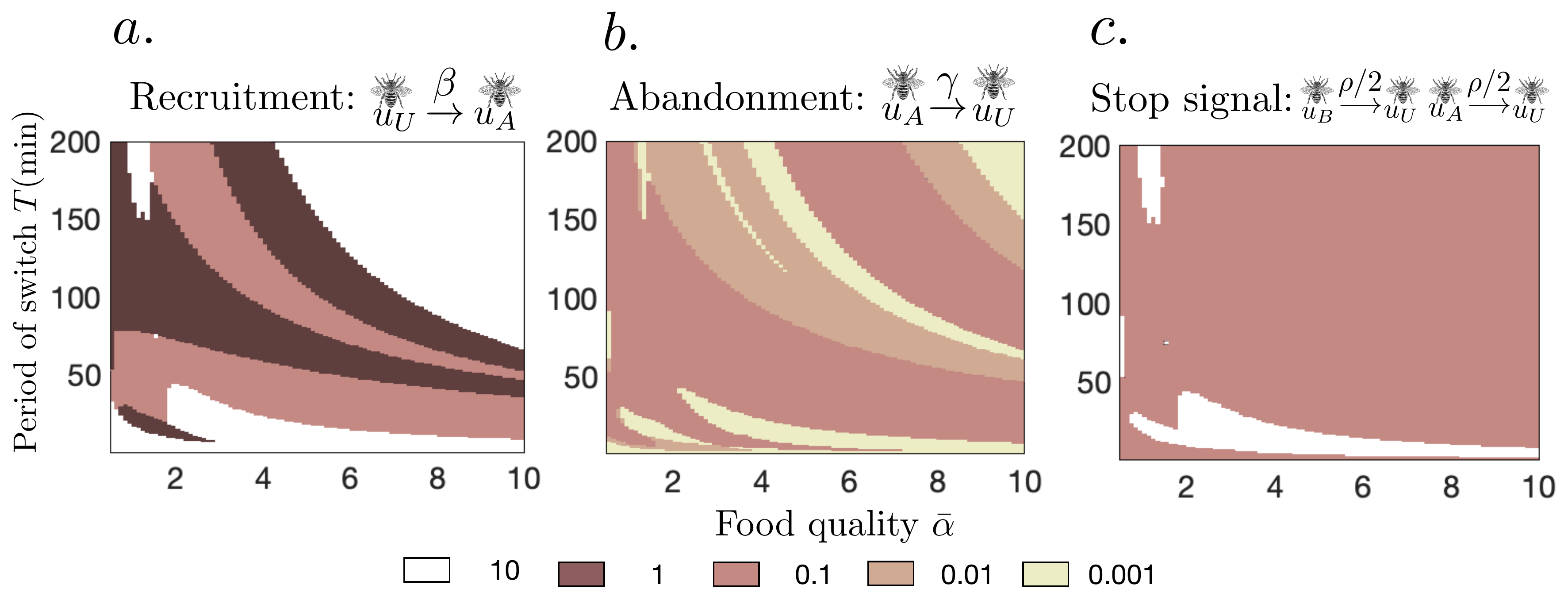}}
\caption{Tuning ({\bf a}) recruitment $\beta$;  ({\bf b}) abandonemnt $\gamma$; and  ({\bf c}) social inhibition $\rho$ to maximize the reward rate (RR) in the indiscriminate stop signaling model. The best tunings of parameters vary considerably with the recruitment being mostly low.}	\label{fig9}
\end{center}
\end{figure}

\subsection{Foraging strategies with discriminate and indiscriminate stop signaling} \label{foragtune}

\begin{figure}[t!]
\vspace{-4mm}
\begin{center}
{\includegraphics[width = 1\linewidth]{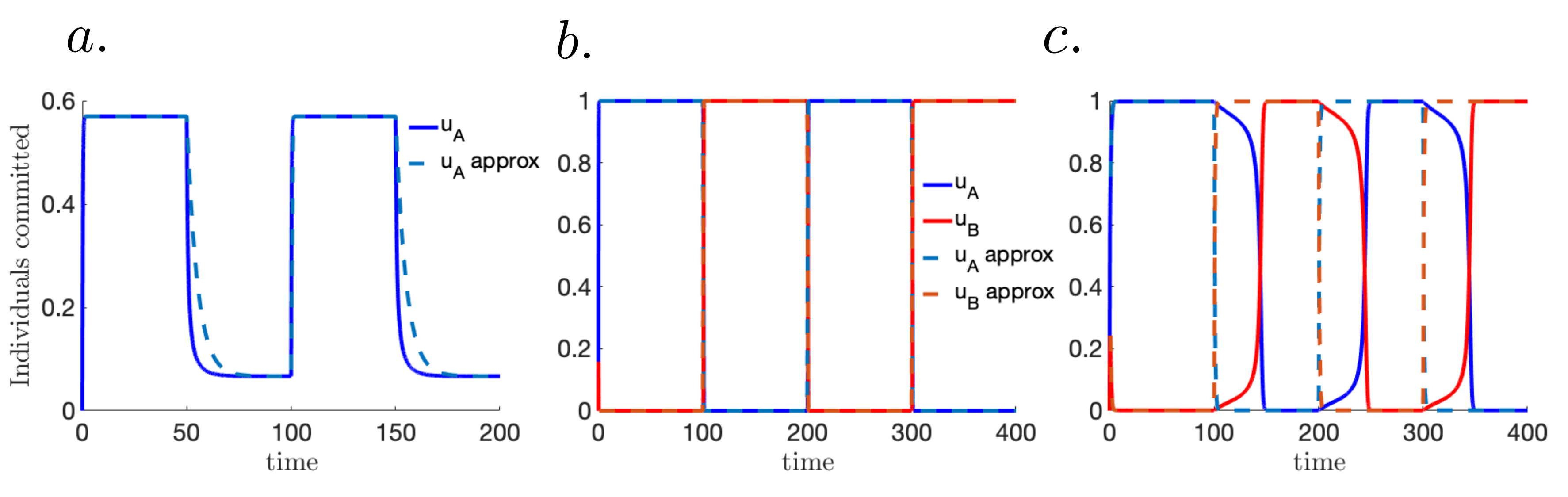}}
\caption{Linear approximation of the switch induced periodic solutions is generally good in ({\bf a}) the single feeder choice model ($\bar{\alpha} = 2$, $T = 50$ min, $\gamma = 2.8$, and $\beta = 3$) and  ({\bf b}) two feeder choice model (direct switching here with model parameters $\bar{\alpha} = 2$, $T = 100$ min, $\gamma = .01$, $\beta = 0.1$ and $\rho = 10$). ({\bf c}) However, when studying the discriminate stop-signaling model close to the saddle-node bifurcation, nonlinear effects shape the periodic solution of the full model Eq.~(\ref{2siteswarm}) in ways that are not well approximated by the linearizations. Model parameters are $\bar{\alpha} = 2$,
$T = 100$ min, $\gamma = .01$, $\beta = 3.6$ and $\rho = 1$.}		\label{fig10}
\end{center}
\end{figure}

\begin{figure}[t!]
\vspace{-4mm}
\begin{center}
{\includegraphics[width = 0.78\linewidth]{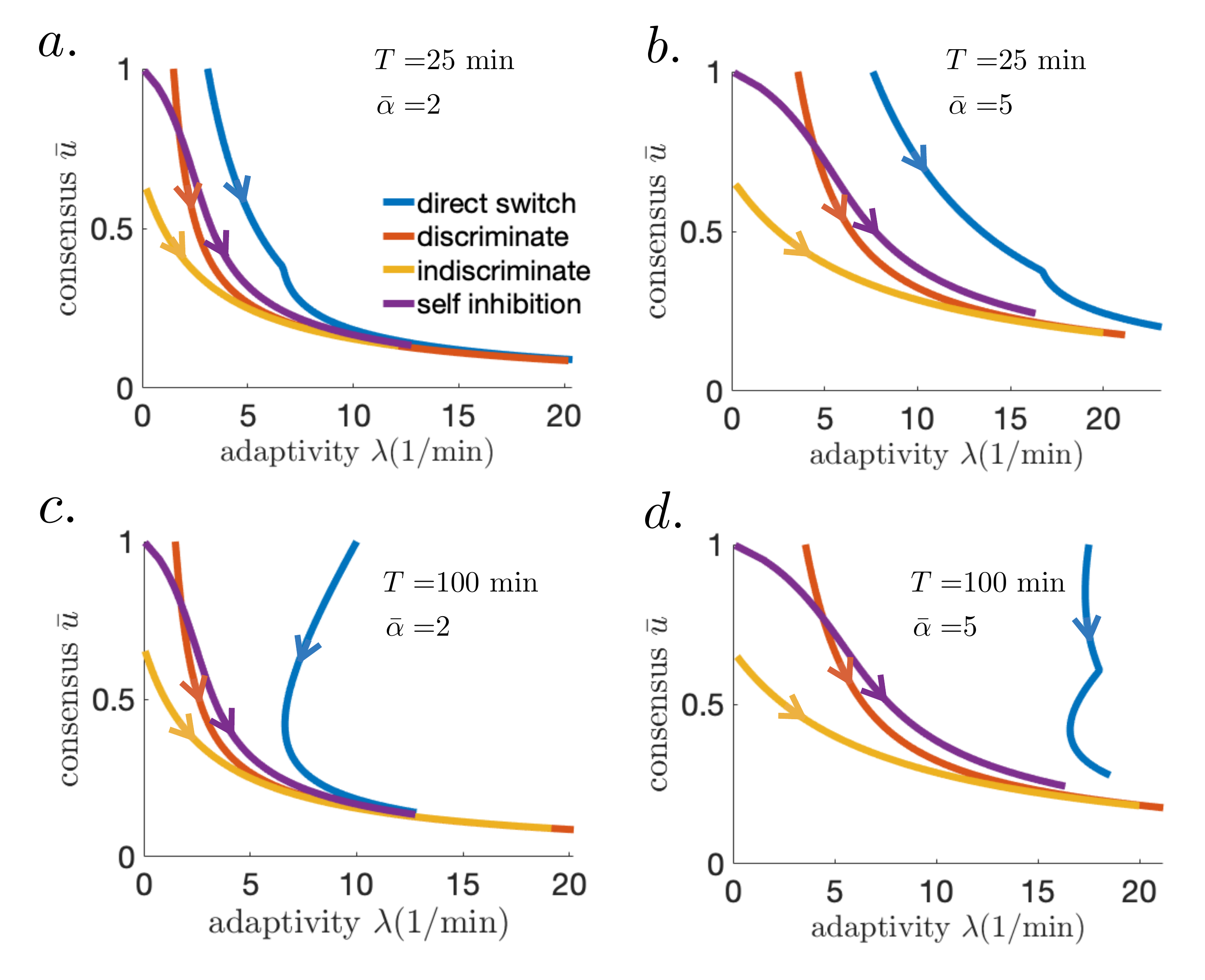}}
\caption{Consensus $\bar{u}$ and adaptivity $\lambda$ computed as described in Sections~\ref{twolinstab} and \ref{linearapptwo}, as abandonment rate $\gamma$ is increased between $[0,20]$ min$^{-1}$ (along the direction of the arrows) for all models. Other parameters are fixed at their optimum level.} \label{fig11} 
\end{center}
\vspace{-6mm}
\end{figure}

Similar to Fig.~\ref{fig5}, we optimized interactions for the discriminate stop signaling and indiscriminate stop signaling model to yield the highest RR Eq.~(\ref{rr2}). 

In discriminate stop signaling model, weak recruitment $\beta$ (Fig.~\ref{fig8}a), weak abandonment $\gamma$ (Fig.~\ref{fig8}b), and strong stop signaling (Fig.~\ref{fig8}c) yield the highest RRs for most environments ($\balpha$, $T$). In slow (large $T$) and high quality $\bar{\alpha}$ environments, abandonment $\gamma$ should be strong, and discriminate stop signaling $\rho$ can be made weak (Fig.~\ref{fig8}b,c). Recruitment should be weak in most environments (Fig.~\ref{fig8}a).

There is no clear preferred interaction profile for maximizing RR across environments ($\balpha$, $T$) in the case of indiscriminate stop signaling (Fig.~\ref{fig9}). Interestingly, the strength of indiscriminate stop signaling parameter $\rho$ should be made low for virtually all environments (Fig.~\ref{fig9}c), and thus it does not seem to improve foraging efficiency. Consensus is lower due to the non-selectivity of social inhibition to all foraging bees.

For the self inhibition model, to maximize foraging efficiency, abandonment should be made weak ($\gamma = 0.01 \text{ min}^{-1}$) while recruitment and social inhibition should be made strong ($\beta = \rho = 10 \text{ min}^{-1}$).

\subsection{Accuracy of linear approximations of periodic solutions}
\label{linac}

Linear approximations of the periodic solutions to the single feeder Eq.~(\ref{singledyn}) and two feeder Eq.~(\ref{2siteswarm}) match the evolution of the full models across a wide range of parameters and forms of social inhibition (for example Fig.~\ref{fig10}a,b). If the system not poised close to a bifurcation, the dynamics between switches roughly linearly decays to the stable equilibrium. However, in the discriminate stop-signaling model, the system can lie close to the saddle-node bifurcation beyond which the model exhibits bistability (Fig.~\ref{fig10}c). In this case, the ghost of the saddle-node slows the solution trajectory, a nonlinear effect which is not well characterized by a linear approximation~\cite{strogatz18}.

\begin{figure}[t!]
\vspace{-9mm}
\begin{center}
{\includegraphics[width = 0.78\linewidth]{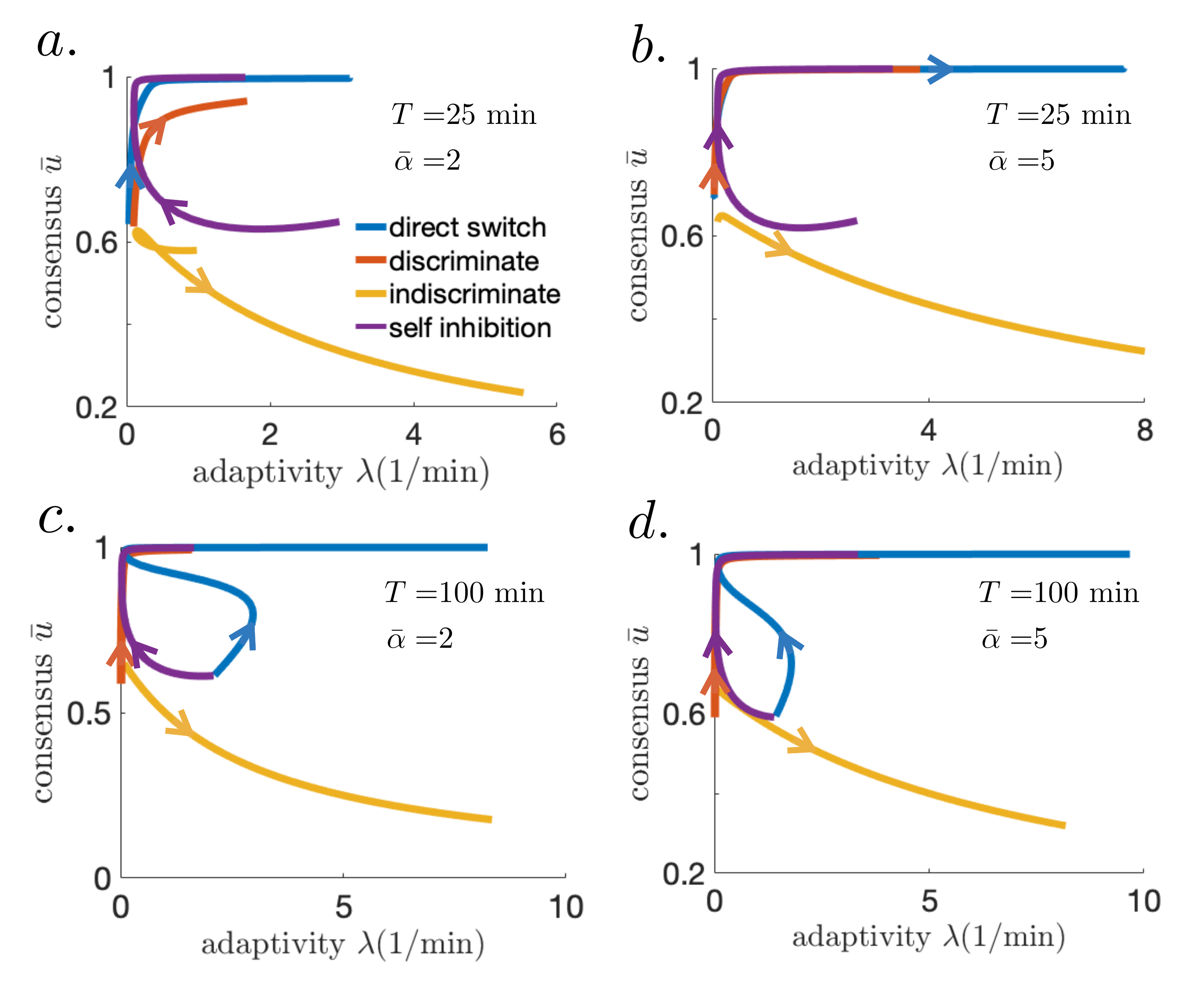}}
\caption{Consensus $\bar{u}$ and adaptivity $\lambda$ computed as described in Sections~\ref{twolinstab} and \ref{linearapptwo}, as social inhibition strength $\rho$ is increased between $[0,20]$ min$^{-1}$ (along the direction of the arrows) for all models. Other parameters are fixed at their optimal level.} \label{fig12} 
\end{center}
\vspace{-6mm}
\end{figure}

\subsection{Computing adaptivity and consensus across models}
\label{adconmod}

Here we calculate consensus $\bar{u}$ and adaptivity $\lambda$ across a wider range of environments as the abandonment rate $\gamma$ (Fig.~\ref{fig11}) and social inhibition rate $\rho$ (Fig.~\ref{fig12}) are varied. The general trends observed in Fig.~\ref{fig6}b,c are preserved. For strong enough abandonment $\gamma$, adaptivity $\lambda$ increases as $\bar{u}$ decreasing, and direct switching tends to balances this trade off best (Fig.~\ref{fig11}). Indiscriminate stop-signaling presents a similar trade off as social inhibition strength $\rho$ is increased (Fig.~\ref{fig12}), while the other social inhibition mechanisms eventually show increases in both consensus $\bar{u}$ and adaptivity $\lambda$, but again direct switching tends to provide higher levels of both overall.

\subsection{Stochastic effects in the finite system size}
\label{stochsys}

\begin{figure}[t!]
\begin{center}
{\includegraphics[width = 0.9\linewidth]{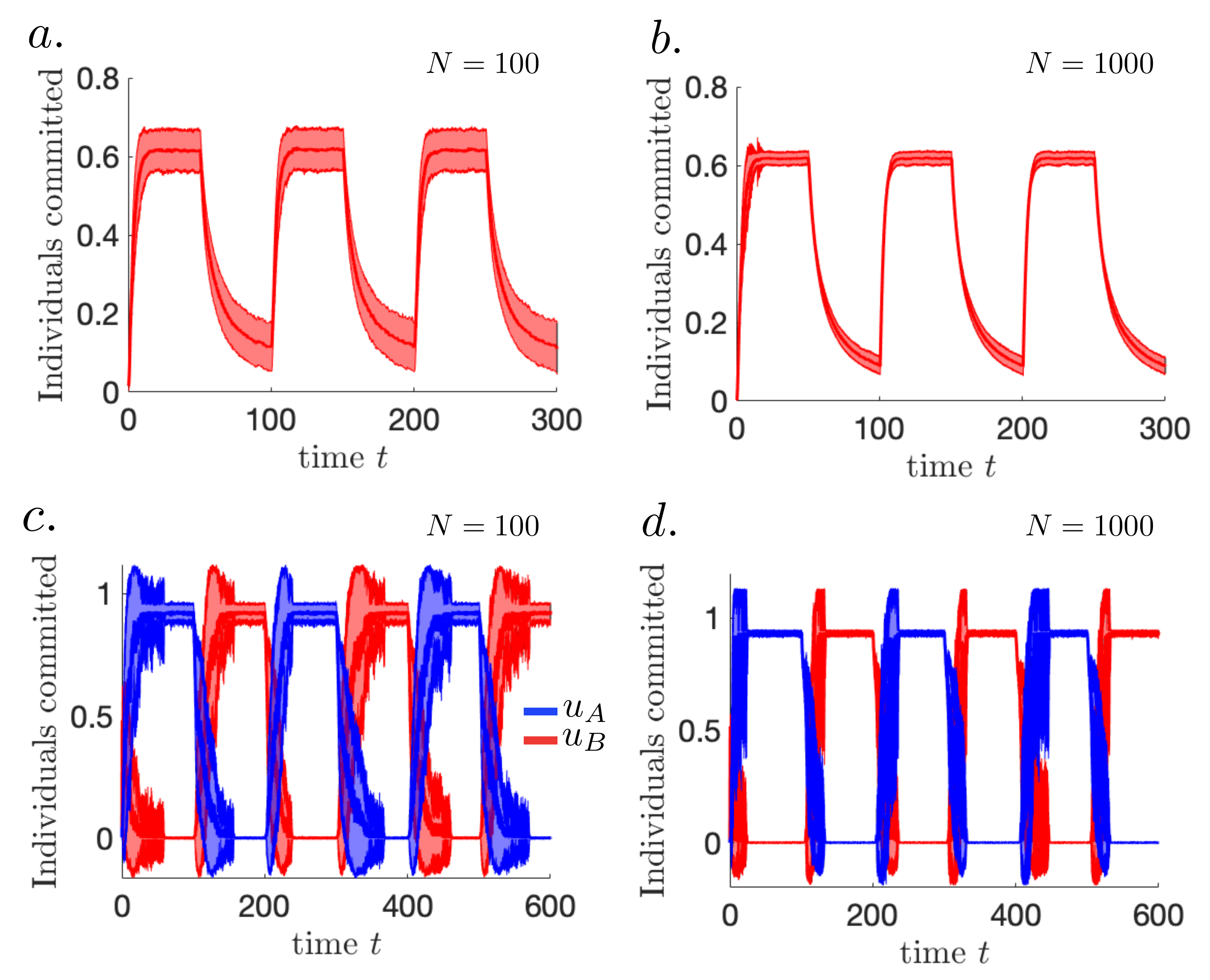}}
\caption{Mean and standard deviation of stochastic simulations of the single feeder Eq.~(\ref{stochone}) and two feeder Eq.~(\ref{mastertwo}) in the case of temporal switching of feeder quality occurring at $T=50$ min intervals. In the single feeder model for ({\bf a}) $N = 100$ and  ({\bf b}) $N = 1000$, near periodic switching of the mean trajectory of simulations (lines) is not far from the behavior of the mean field system Eq.~(\ref{singledyn}). Other model parameters are $\tilde{\alpha} = 0.2$, $\tilde{\beta} = 0.2$,  and $\tilde{\gamma} = 0.2$. Standard deviations (shaded regions) decrease as the system size is increased. Similar trends are apparent in the statistics of the two feeder finite-sized models with discriminate stop signaling  ({\bf c}) $N = 100$ and ({\bf d}) $N = 1000$. Other model parameters are $\tilde{\alpha} = 2$, $\tilde{\beta} = 2$, $\tilde{\gamma} = 0.1$, $\tilde{\rho}= 1$, $T=100$ min and $\tau = 0$.} 	\label{fig13}
\end{center}
\end{figure}

Honeybee swarms tend to be of modest size (in the 1,000s)~\cite{seeley10}, and so it is reasonable to expect some impact of finite size effect on the dynamics of foraging. In general, we found finite size effects induced fluctuations about the typical mean periodic switching solutions, but that it did not qualitatively alter the general behavior of the swarm (Fig.~\ref{fig13}). The finite size model is governed by a master equation, determining the probability of all possible changes in committed and uncommitted populations.

In the case of the single feeder model, the master equation for the probability $p(n,t)$ of finding $n$ bees committed to foraging at time $t$ is
\begin{align}
\dot{p}(n,t) = r_+(n-1) p(n-1,t) + r_-(n+1) p(n+1,t) - \left[ r_+(n) + r_-(n) \right] p(n,t), \label{stochone}
\end{align}
for integer $n=0,1,2,...,N$ with boundary conditions $p(-1,t) = p(N+1,t) = 0$ and forward and backward transition rates
\begin{align*}
r_+(n) = (N-n) ( \tilde{\alpha} (t) + \tilde{\beta} n ), \hspace{5mm} r_-(n) = \tilde{\gamma} n + \tilde{\rho}( \bar{\tilde{\alpha}} - \tilde{\alpha}(t - \tau)) n^2
\end{align*}
for system size (total bee number) $N$. To obtain the mean field Eq.~(\ref{singledyn}) as $N \to \infty$~\cite{seeley12,vankampen92}, one must define $\tilde{\alpha}(t) = \alpha (t) /N$, $\tilde{\beta} = \beta / N^2$, $\tilde{\gamma} = \gamma / N$, and $\tilde{\rho} = \rho/N^2$. Note, the scalings correspond to the power of the population count appearing in the interaction term, ensuring the transition terms remain bounded in the thermodynamic limit. We utilized the stochastic simulation algorithm by Gillespie~\cite{gillespie77} to evolve the stochastic system for the statistic plotted in Fig.~\ref{fig13}a,b. We make two remarks about our findings. First, the swarm generally increases the fraction of committed foragers when food is present at the feeder and decreases when food is removed. Second, the amplitude of fluctuations in individual simulations decreases with system size, as typically expected~\cite{vankampen92}, as evidenced by the narrower standard deviations in the solution trajectories in the $N=1000$ versus the $N=100$ simulations.

In the case of the two feeder model, the master equation is more complicated as it must track the probability of transitions between uncommitted bees, bees committed to feeder A, and those committed to feeder B. Indeed, we can write the model down for any of the four forms of social inhibition, but we just provide the discriminate stop signaling model here. Others can be written similarly. The probability $p(n_A, n_B, t)$ of finding $n_A$ bees committed to A and $n_B$ committed to $B$ at time $t$ given system size $N$ is (dropping the argument in $t$ for brevity):
\begin{align}
\dot{p}(n_A, n_B) =& r_{0A}(n_A-1,n_B)p(n_A-1,n_B) + r_{0B}(n_A, n_B-1) + r_{A0}(n_A+1,n_B) \label{mastertwo} \\
& + r_{B0}(n_A, n_B+1) p(n_A, n_B+1)  - \left[ r_{0A}(n_A, n_B) + r_{0B}(n_A, n_B) + \right. \nonumber \\
& \left. + r_{A0}(n_A,n_B) + r_{B0}(n_A, n_B) \right]p(n_A, n_B), \nonumber
\end{align}
for $n_A, n_B = 0, 1, ..., N$ with the condition that $n_A + n_B \leq N$, boundary conditions $p(-1, n_B) = p(n_A, -1) = p(N+1, n_B) = p(n_A, N+1) = 0$, and transition rates
\begin{align*}
r_{0A}(n_A, n_B) &= (N-n_A - n_B) ( \tilde{\alpha}_A(t) + \tilde{\beta} n_A), \hspace{6mm} r_{0B}(n_A, n_B) = (N-n_A - n_B) ( \tilde{\alpha_A(t)} + \tilde{\beta} n_B), \\
r_{A0}(n_A, n_B) &= \tilde{\gamma} n_A + \tilde{\rho} \alpha_B(t- \tau)n_A n_B, \hspace{15mm} r_{B0}(n_A, n_B) = \tilde{\gamma} n_B + \tilde{\rho} \alpha_A(t- \tau)n_A n_B.
\end{align*}
As in the single feeder model, periodic switching with environmental switches is apparent, and the amplitude of fluctuations decreases with system size (Fig.~\ref{fig13}c,d).

A detailed study of the finite size population model would require a much more thorough treatment and statistical analysis. We expect the effects of stochasticity will not considerably impact our general findings. The only qualitative differences we would expect would be in the case of unrealistically small systems (e.g., $N=10$), and in bistable systems (like cases of the discriminate stop signaling model), where fluctuations could drive switching between multiple stable equilibria~\cite{biancalani14}.

\end{document}